\documentclass[prb,superscriptaddress,showpacs,floatfix,tightenlines,10pt,twocolumn]{revtex4-1}
\usepackage{amssymb}
\usepackage{amsmath}
\usepackage{graphicx}
\usepackage{float}

\newcommand{\func}[1]{\operatorname{#1}}

\begin{document}

\title{Magneto-optical Kerr effect and signature of the chiral anomaly in a
Weyl semimetal in a magnetic field}
\author{Jean-Michel Parent}
\author{Ren\'{e} C\^{o}t\'{e}}
\author{Ion Garate}
\affiliation{D\'{e}partement de physique and Institut Quantique, Universit\'{e} de
Sherbrooke, Sherbrooke, Qu\'{e}bec, J1K 2R1, Canada }
\date{\today }

\begin{abstract}
One striking property of the Landau level spectrum of a Weyl semimetal (WSM)
is the existence of a chiral Landau level, in which the electrons propagate
unidirectionally along the magnetic field. This linearly dispersive level
influences the optical properties of WSMs. For example, it was recently
shown that a complete optical valley polarization is achievable in a
time-reversal symmetric Weyl semimetal placed in a magnetic field\cite%
{Bertrand2019}. This effect originates from inter-Landau level transitions
involving the chiral Landau level and requires a tilt of the Weyl cones. In
this paper, we show how the magneto-optical Kerr effect (MOKE) is modified
in a WSM\ with tilted Weyl cones in comparison with its behavior in a normal
metal and how a valley polarization can be detected using MOKE. We study
both the Faraday (longitudinal) and Voigt (transverse) configurations for
light incident on a semi-infinite WSM surface with no Fermi arcs. We use a
minimal model of a WSM with four tilted Weyl nodes related by mirror and
time-reversal symmetry. In the Voigt configuration, a large peak of the Kerr
angle occurs at the plasmon frequency. We show that the blueshift in
frequency of this peak with increasing magnetic field is a signature of the
chiral anomaly in the MOKE.
\end{abstract}

\maketitle

\section{INTRODUCTION}

A Weyl semimetal\cite{WSM review} is a three-dimensional topological phase
of matter, where pairs of nondegenerate bands cross at isolated points in
the Brillouin zone. Near these points, called \textquotedblleft Weyl
nodes\textquotedblright , the electronic dispersion is gapless and linear in
momentum and the excitations satisfy the Weyl equation, a two-component
analog of the Dirac equation. Each Weyl node is a source or sink of Berry
curvature, which acts as a magnetic field in momentum space, and has a
chirality index $\chi =\pm 1$ reflecting the topological nature of the band
structure. The Nielsen-Ninomiya theorem\cite{Nielsen} requires that the
number of Weyl points in the Brillouin zone be even so that Weyl nodes must
occur in pairs of opposite chirality. For the Weyl nodes to be stable,
either inversion symmetry or time-reversal symmetry must be broken.

Weyl semimetals show a number of interesting transport properties, such as
an anomalous Hall effect\cite{AHE}, a chiral-magnetic effect\cite{CME},
Fermi arcs\cite{Fermi arcs} and a chiral anomaly leading to a negative
longitudinal magnetoresistance\cite{Chiral anomaly}. The topological aspects
of WSMs also show up in their optical properties, especially so when a
magnetic field is present. In this case, the linear dispersion is split into
positive ($n>0$) and negative ($n<0$) energy dispersive Landau levels. The $%
n=0$ Landau level is chiral because its dispersion$\ $is unidirectional,
e.g. $E\left( k_{z}\right) =-\chi v_{F}k_{z}$ for a magnetic field along the 
$z$ direction. The absorption spectrum is different from that of Schr\"{o}%
dinger or Dirac fermions\cite{MagnetoSigma} and can be used to show the
phenomenon of charge pumping due to the chiral anomaly\cite{CarbotteCA} or
other photoinduced responses, as well as to distinguish between type I and
type II\ WSMs\cite{Goerbig2016}.

In this paper, we investigate another optical property that is affected by
the topological nature of WSMs, i.e. the magneto-optical Kerr effect (MOKE),
which consists in the rotation of the plane of polarization of a beam of
light reflected from the surface of a WSM in a magnetic field. In graphene,
also a material with Dirac-like dispersion, a substantial rotation of the
polarization plane ($>0.1$ rad) upon transmission (the related Faraday
effect) has been reported recently\cite{Faraday graphene}. In WSMs, Faraday
and Kerr rotations have been studied in some detail in Ref. 
\onlinecite{Kerr
Randeria} for a minimal model of a WSM with intrinsically broken
time-reversal symmetry (TRS) and no magnetic field. In such model, the axion
term of the electromagnetic action makes a gyrotropic contribution to the
dielectric function, thereby leading to Faraday and Kerr rotations in the
absence of an external magnetic field.

In the present work, our model of a WSM\ preserves TRS and the Kerr rotation
is due to the presence of an external magnetic field, which we set either
along the direction of propagation of the incoming electromagnetic wave
(i.e. the longitudinal or Faraday configuration) or perpendicular to it (the
transverse or Voigt configuration). One motivation for this work is our
previous study of the optical absorption\cite{Bertrand2019,Bertrand2017} in
WSMs, which predicted the possibility of a complete optical valley
polarization for a sizeable interval of frequency in a time-reversal
symmetric type I WSM with tilted Dirac cones, by a suitable choice of the
relative orientation of the incoming light wave, magnetic field, and tilt
vector. The valley polarization shows up as a splitting of the absorption
line of two nodes related by TRS at zero magnetic field, for transitions
involving the chiral Landau level.

There have been some previous works on the MOKE\ in WSMs. A giant
polarization rotation has been predicted in type I\ and II WSM\ with tilted
cones and broken TRS in zero magnetic field\cite{Sonowal}. Kerr and Faraday
rotations for zero tilt but finite magnetic field and broken TRS have also
been studied\cite{JYang}. Moreover, experimental evidences for chiral
pumping of the Weyl nodes in the WSM TaAs have appeared recently\cite%
{Levy,Cheng}. The work we present here is different. We study the MOKE\ in a
simplified model of a WSM with four nodes related by TRS and mirror symmetry
in a quantizing magnetic field and in both the Faraday and Voigt geometries.
We show that, in contrast with a \textquotedblleft normal\textquotedblright\
metal, in a WSM a sizeable Kerr rotation can be expected in both geometries
for moderate values of the background dielectric constant $\varepsilon
_{\infty }$. In the resonant regime, where the frequency of the incoming
light matches an electronic interband (i.e. inter-Landau level) transition,
Kerr rotation can be used as a spectroscopic tool to detect the inter-Landau
level transitions. The presence of tilted cones modifies the Landau level
quantization and changes the selection rules, giving a much richer interband
spectrum in MOKE: when a magnetic field is applied in a direction other than
the tilt, interband transitions other than the usual dipolar ones ($%
\left\vert n\right\vert \rightarrow \left\vert n\right\vert \pm 1$) become
possible\cite{Goerbig2016}. Moreover, the valley polarization effect we
reported earlier for optical absorption also appears in the Kerr rotation,
thus providing another way to detect this effect experimentally.

We find that the Voigt configuration is particularly interesting because it
enables having a component of the incoming electric field in the direction
of the quantizing magnetic field. A consequence of the chiral anomaly in
WSMs is that the plasmon frequency $\omega _{p},$ which is given by the
condition that $\func{Re}\left[ \varepsilon _{\Vert }\left( \omega
_{p}\right) \right] =0,$ increases with magnetic field. Here $\varepsilon
_{\Vert }$ is the element of the dielectric tensor in the direction of the
external magnetic field, e.g. $\varepsilon _{xx}$ for $\mathbf{B}$ along $%
\widehat{\mathbf{x}}.$ In contrast with the Faraday configuration, $%
\varepsilon _{\Vert }$ enters in the definition of the Kerr angle so that we
expect that the behavior of the Kerr angle will be modified by the chiral
anomaly. Indeed, we show that a strong maximum in the Kerr angle occurs at
the plasmon frequency, which is in the THz range for moderate values of $%
\varepsilon _{\infty },$ i.e. close to the threshold of the electronic
interband transitions. The frequency of this peak increases with magnetic
field, providing a clear signature of the chiral anomaly in the Kerr
rotation.

The remainder of this paper is organized as follows:\ Section II\ introduces
our minimal four-node model of a WSM with TRS and tilted cones, and gives
the energy spectrum of each node. Section III explains how we compute the
dynamical conductivity tensor for both inter- and intra-Landau level
transitions. In Sec. IV, we give the formalism to compute the Kerr angle in
both the Faraday and Voigt configurations. Section V contains our numerical
results, which are further summarized in Sec. VI. In order to lighten the
main text, we have put details of all calculations in appendix A for the
energy spectrum and appendix B for the derivation of the current operator
for tilted cones. In appendix C, we discuss the MOKE\ for a normal metal in
order to provide a basis for comparison with our findings for a WSM.

\section{MODEL HAMILTONIAN}

We consider a simple model of a WSM, which possesses TRS in the absence of a
magnetic field. This toy model has been described and justified in Refs. %
\onlinecite{Bertrand2017,Bertrand2019}, where we used it to calculate the
optical valley polarization in a WSM. The model consists of four tilted Weyl
nodes (denoted by the index $\tau =1,2,3,4$), two for each chirality, and a
mirror plane placed perpendicularly to the $z$ axis as shown in Fig. \ref%
{fig1}. Pairs of nodes of opposite chirality ($\tau =1,2$ and $\tau =3,4$)
are related to one another by the mirror plane, while nodes $\tau =1,3$ and $%
\tau =2,4$ are related by time-reversal symmetry in the absence of the
magnetic field. Thus, the four nodes are symmetry-equivalent in the absence
of a magnetic field.

\begin{figure}
\centering
\includegraphics[width = \linewidth]{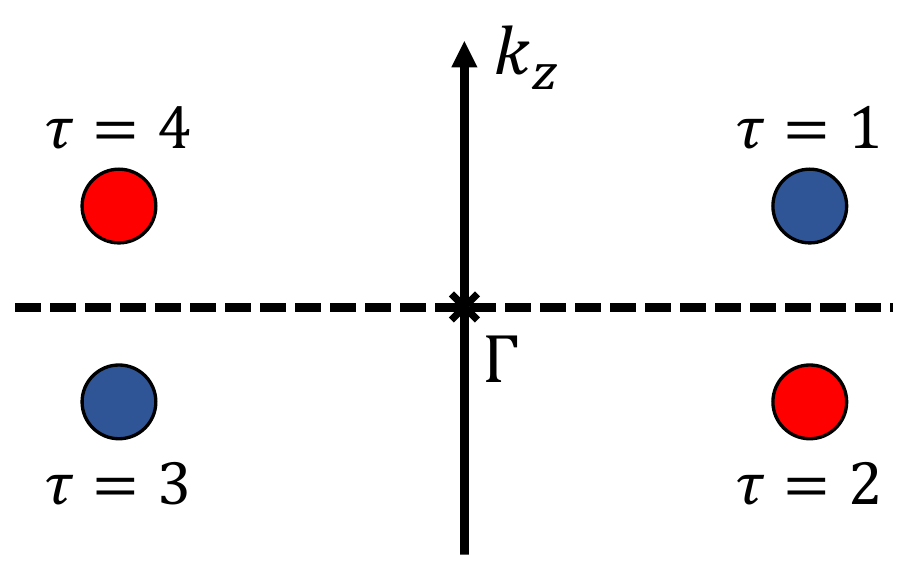} 
\caption{Toy model of a WSM with
		time-reversal symmetry and a mirror plane perpendicular to the $\protect\widehat{\mathbf{z}}$ 
		direction. The $y$ axis passes through the cross. Blue
		and red circles indicate Weyl nodes with opposite chiralities.} \label%
{fig1}
\end{figure}

The low-energy noninteracting single-particle Hamiltonian for an electron in
node $\tau $ is given by%
\begin{equation}
h_{\tau }(\mathbf{p})=d_{\tau ,0}(\mathbf{p})\sigma _{0}+\mathbf{d}_{\tau }(%
\mathbf{p})\cdot \mathbf{\sigma ,}  \label{htau}
\end{equation}%
where $\mathbf{p}$ is the momentum of the electron measured with respect to
the Weyl node, $\mathbf{\sigma }$ is a vector of Pauli matrices in the $1/2-$%
pseudospin state of the two bands at their crossing points and $\sigma _{0}$
is the $2\times 2$ unit matrix. For node $\tau =1,$ we take%
\begin{align}
d_{1,0}(\mathbf{p})& =v_{F}\mathbf{t\cdot p},  \notag \\
\mathbf{d}_{1}(\mathbf{p})& =v_{F}\mathbf{p},
\end{align}%
where $v_{F}$ is the Fermi velocity and $\mathbf{t}$ is a dimensionless
vector describing the magnitude and direction of the tilt of the Weyl cone.
We restrict our analysis to a type I WSM, i.e. to $\left\vert \mathbf{t}%
\right\vert <1$. The Hamiltonians of the other three Weyl nodes are obtained
by applying mirror and time-reversal operations to $h_{1}(\mathbf{p})$.
These amount to making the transformations%
\begin{align}
\tau & =1\rightarrow 2:(v_{F},t_{x},t_{y},t_{z})\rightarrow
(-v_{F},-t_{x},-t_{y},t_{z}),  \notag \\
\tau & =1\rightarrow 3:(v_{F},t_{x},t_{y},t_{z})\rightarrow
(v_{F},-t_{x},-t_{y},-t_{z}),  \label{transfo2} \\
\tau & =1\rightarrow 4:(v_{F},t_{x},t_{y},t_{z})\rightarrow
(-v_{F},t_{x},t_{y},-t_{z}),  \notag
\end{align}%
where we have assumed that $\boldsymbol{\sigma }$ transforms as a spin under
time-reversal and mirror operations (see Ref. \onlinecite{Bertrand2017} for
a discussion of this point).

A transverse static magnetic field $\mathbf{B}_{0}$ is added via the Peierls
substitution $\mathbf{p}\rightarrow \mathbf{P}=\mathbf{p}+e\mathbf{A.}$
While it is possible to obtain the energy levels of the hamiltonian
analytically\cite{Goerbig2016}, we find it more convenient to use a
numerical approach in order to get a fully orthonormal basis for the
eigenspinors. Moreover, a numerical approach allows a study of the system
for arbitrary orientations of the magnetic field and tilt vector. It also
allows the consideration of more complex Hamiltonians with, for example, non
linear terms in the energy spectrum\cite{Bertrand2017}. The numerical
diagonalization of $h_{1}(\mathbf{P})$ is carried out in Appendix A. The
many-body Hamiltonian in the basis of the Landau levels of $h_{1}(\mathbf{P}%
) $ can be written as

\begin{equation}
\mathcal{H}=\sum_{\tau ,X,p_{\Vert },I}E_{I}\left( \tau ,p_{\Vert }\right)
d_{I}^{\dag }\left( \tau ,X,p_{\Vert }\right) d_{I}\left( \tau ,X,p_{\Vert
}\right) .  \label{hamiltone}
\end{equation}%
In Eq. (\ref{hamiltone}), the eigenstates are defined by the set of quantum
numbers $\left( I,X,p_{\Vert },\tau \right) ,$ where $X$ is the
guiding-center index and $I=\left( n,s\right) $ the Landau level index. Our
convention is to take $n=0,1,2,3,...$ as a positive number and use $%
s=+1\left( -1\right) $ for the positive (negative)-energy levels$.$ The
variable $p_{\Vert }=\hslash k_{\Vert }$ is the momentum in the direction of
the magnetic field, which we set along the $z$ (or $x$) axis, i.e.
perpendicular (or parallel) to the mirror plane. Each Landau level $\left(
I,p_{\Vert },\tau \right) $ has the macroscopic degeneracy $N_{\varphi
}=S/2\pi \ell ^{2},$ where $\ell =\sqrt{\hslash /eB_{0}}$ is the magnetic
length and $S$ is the area of the WSM\ perpendicular to the magnetic field.
The operator $d_{I}^{\dag }\left( \tau ,X,p_{\Vert }\right) $ creates an
electron in the quantum state $\left( I,p_{\Vert },X,\tau \right) .$

\begin{figure}
\centering
\includegraphics[width = \linewidth]{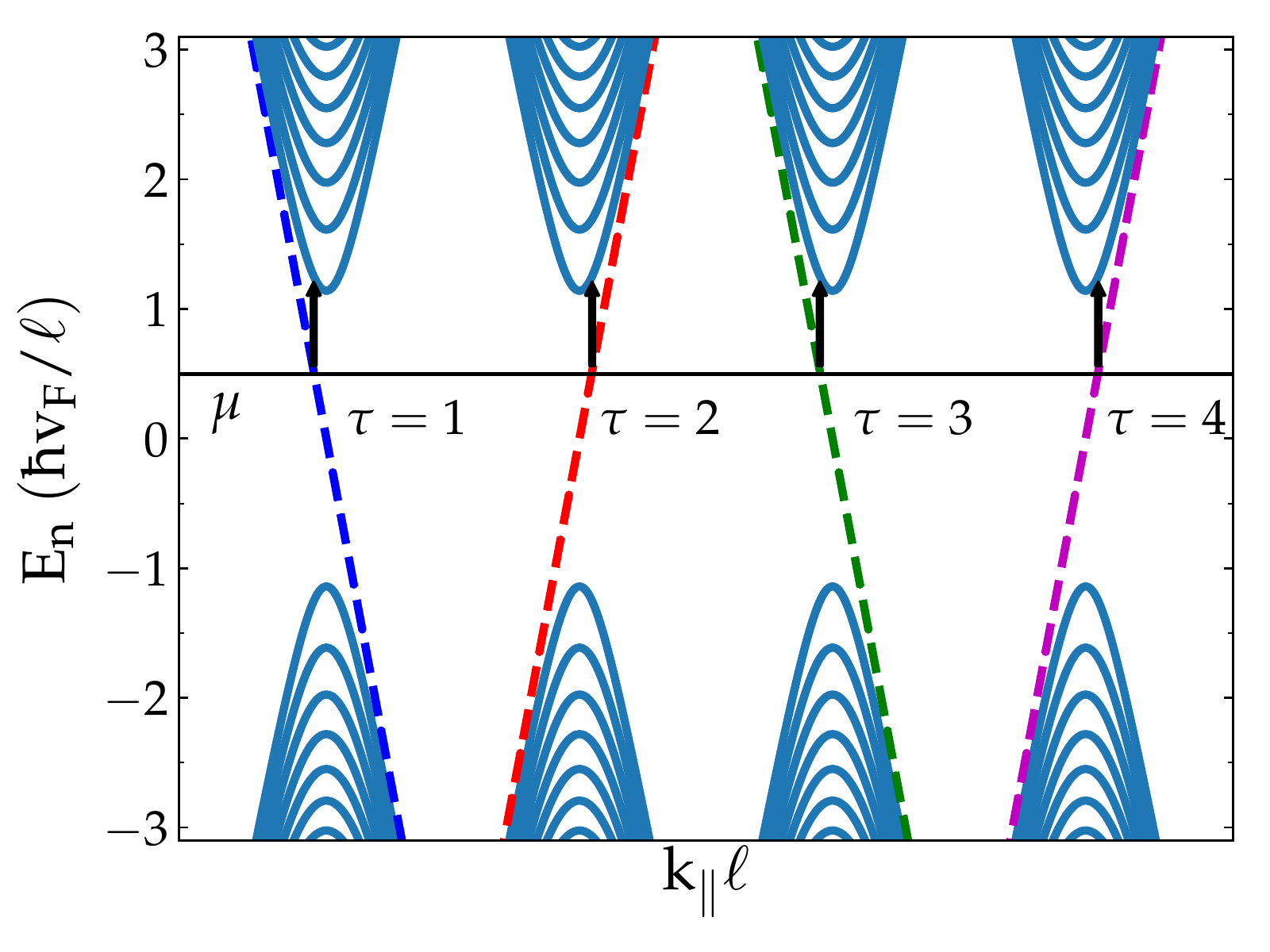} 
\caption{Dispersion of the first Landau
		levels of the four nodes for $\mathbf{t}=t\protect\widehat{\mathbf{z}}$ and 
		$\mathbf{B}=B_{0}\protect\widehat{\mathbf{x}}.$ The length of each arrow
		indicates the lowest-energy interband transition under right circularly
		polarized light. When the tilt vectors of the nodes are perpendicular to the static magnetic field, all nodes have the same interband absorption threshold. For clarity, the position of the different nodes has been
		shifted in $k_{\Vert }\ell .$} \label{fig2}
\end{figure}

The dispersion of the Landau levels is given in Figs. \ref{fig2} and \ref%
{fig3} for $\mathbf{t}\perp\mathbf{B}$ and $\mathbf{t}||\mathbf{B}$,
respectively. The arrows in these figures indicate the energy gap for
interband transitions when the Fermi level $E_{F}$ is in the chiral Landau
level (i.e. $n=0$). The optical gaps are identical for all nodes when the
tilt and magnetic field are perpendicular to each other. When the magnetic
field is perpendicular to the mirror plane and has a nonzero projection
along the tilt vector, the optical gaps for interband transitions involving
the chiral Landau level are different for nodes that are related by
time-reversal symmetry when $\textbf{B}_{0}=0$, but equal for nodes that are
mirror partners (see Ref. \onlinecite{Bertrand2019} for a discussion on
general orientations of $\textbf{B}$). When light propagates along the
direction of the magnetic field, this difference in the absorption gap leads
to a full valley polarization\cite{Bertrand2019}.

\begin{figure}
\centering
\includegraphics[width = \linewidth]{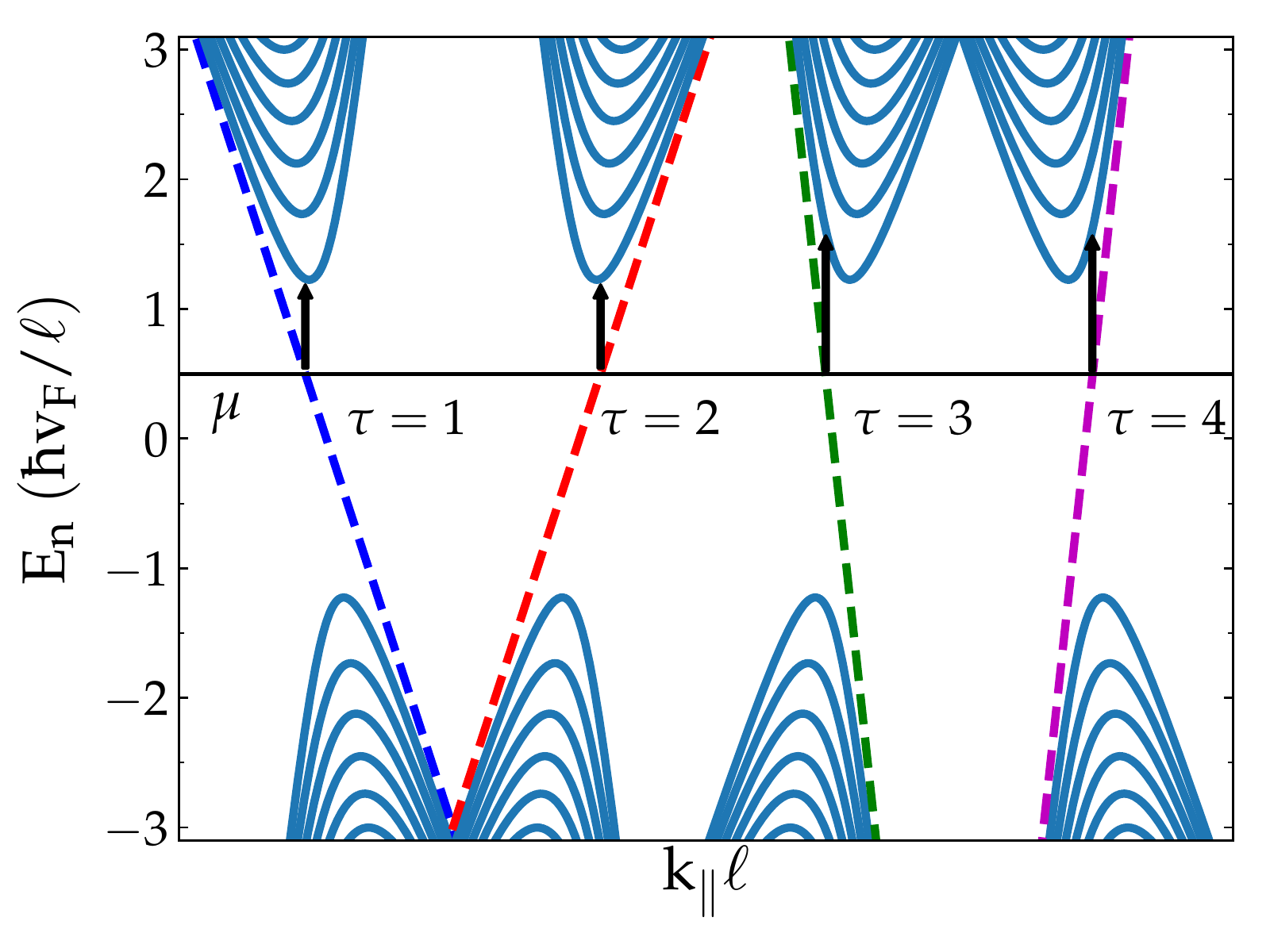} 
\caption{Dispersion of the Landau levels
		for the four nodes for $\mathbf{t}=t\protect\widehat{\mathbf{z}}$ and 
		$\mathbf{B}=B_{0}\protect\widehat{\mathbf{z}}$. The length of each arrow
		indicates the lowest-energy interband transition under right circularly
		polarized light. When the tilt vectors of the nodes have nonzero components along the static magnetic field, different pairs of nodes have different interband absorption thresholds in the quantum limit, thereby leading to a valley polarization. For clarity, the position of the different nodes has been
		shifted in $k_{\Vert }\ell .$} \label{fig3}
\end{figure}

\section{OPTICAL CONDUCTIVITY}

The magneto-optical conductivity of Weyl semimetals has been calculated
before\cite{Ashby}. The effect of a tilt on the ac optical response has also
been considered in a model of a WSM\ with broken TRS and no quantizing
magnetic field\cite{mukherjee}. Here we consider the effect of a tilt on the
four-node model introduced above (which has TRS) and in the presence of a
quantizing magnetic field. We restrict ourselves to $T=0$K. Since we are
mainly interested in the behavior of the Kerr angle in the resonant regime,
i.e. in the THz range of frequencies, finite temperature effects should be
negligible for experiments carried out at low temperature, i.e. at a few
Kelvin.

The single-particle current operator for node $\tau $is defined by%
\begin{equation}
\mathbf{j}_{\tau }=-\left. \frac{\partial h_{\tau }(\mathbf{p})}{\partial 
\mathbf{A}_{ext}}\right\vert _{\mathbf{A}_{ext}\rightarrow 0}=-ev_{F}\left( 
\mathbf{\sigma }+\mathbf{t}\sigma _{0}\right) ,
\end{equation}%
where $\mathbf{A}_{ext}$ is the vector potential of an external
electromagnetic field. The many-body current is 
\begin{eqnarray}
\mathbf{J}_{\tau } &=&\int d^{3}r\psi _{\tau }^{\dag }\left( \mathbf{r}%
\right) \mathbf{j}_{\tau }\psi _{\tau }\left( \mathbf{r}\right) \\
&=&-ev_{F}\sum_{\tau ,p_{\Vert },X}\sum_{I,J}\mathbf{\Upsilon }_{I,J}\left(
\tau ,p_{\Vert }\right) d_{I}^{\dag }\left( \tau ,X,p_{\Vert }\right)
d_{J}\left( \tau ,X,p_{\Vert }\right) ,  \notag
\end{eqnarray}%
where the matrix elements $\mathbf{\Upsilon }_{I,J}\left( \tau ,p_{\Vert
}\right) $ are given in Eq. (\ref{matrixe}) of Appendix B.

It is convenient to define the operator 
\begin{equation}
\rho _{I,J}\left( \tau ,p_{\Vert }\right) =\frac{1}{N_{\varphi }}%
\sum_{X}d_{I}^{\dag }\left( \tau ,X,p_{\Vert }\right) d_{J}\left( \tau
,X,p_{\Vert }\right) ,
\end{equation}%
so that the many-body Hamiltonian and current can be written as%
\begin{equation}
\mathcal{H}=\sum_{\tau ,p_{\Vert },I}E_{I}\left( \tau ,p_{\Vert }\right)
\rho _{I,I}\left( \tau ,p_{\Vert }\right)
\end{equation}%
and 
\begin{equation}
\mathbf{J}_{\tau }=-ev_{F}\sum_{p_{\Vert }}\sum_{I,J}\mathbf{\Upsilon }%
_{I,J}\left( \tau ,p_{\Vert }\right) \rho _{I,J}\left( \tau ,p_{\Vert
}\right) .
\end{equation}

The optical conductivity tensor is related to the retarded current response
function $\chi _{\alpha ,\beta }^{R}(\omega )=\chi _{\alpha \beta }^{R}(%
\mathbf{q}=0,\omega )$ by 
\begin{equation}
\sigma _{\alpha ,\beta }(\omega )=\frac{i}{\omega }\left[ \chi _{\alpha
,\beta }^{R}(\omega )-\chi _{\alpha ,\beta }^{R}(0)\delta _{\alpha \beta }%
\right] ,  \label{sigmas}
\end{equation}%
where $\alpha ,\beta \in \left\{ x,y,z\right\} $ and $\omega $ is the
frequency of the incoming light beam. The diamagnetic contribution to the
current operator is lacking in the continuum approximation of the linear
spectrum of the low-energy model. This absence leads to unphysical terms\cite%
{Kerr Randeria} in $\sigma _{\alpha ,\beta }(\omega )$. The second term on
the right hand side of Eq. (\ref{sigmas}) is required in order to remove
these spurious contributions.

The retarded current response function can be obtained from the two-particle
Matsubara Green's function 
\begin{eqnarray}
\chi _{\alpha ,\beta }\left( \tau \right) &=&-\frac{1}{\hslash V}%
\left\langle T_{\tau }J_{\alpha }\left( \tau \right) J_{\beta }\left(
0\right) \right\rangle \\
&=&-e^{2}v_{F}^{2}\frac{1}{\hslash V}\sum_{I,J}\sum_{K,L}\sum_{p_{\Vert
},p_{\Vert }^{\prime }}\Upsilon _{I,J}^{\left( \alpha \right) }\left(
p_{\Vert }\right) \Upsilon _{K,L}^{\left( \beta \right) }\left( p_{\Vert
}^{\prime }\right)  \notag \\
&&\times \left\langle T_{\tau }\rho _{I,J}\left( \tau ,p_{\Vert }\right)
\rho _{K,L}\left( 0,p_{\Vert }^{\prime }\right) \right\rangle ,  \notag
\end{eqnarray}%
where $T_{\tau }$ is the time ordering operator and $\tau $ an imaginary
time, not to be confused with the node index. We omit the node index in the
remaining of this section in order to avoid any confusion. In linear
response, $\chi _{\alpha ,\beta }\left( \tau \right) $ is approximated by%
\begin{eqnarray}
\chi _{\alpha ,\beta }\left( \tau \right) &=&e^{2}v_{F}^{2}\frac{1}{\hslash V%
}\sum_{I,J}\sum_{X,p_{\Vert }}\Upsilon _{I,J}^{\left( \alpha \right) }\left(
p_{\Vert }\right) \Upsilon _{J,I}^{\left( \beta \right) }\left( p_{\Vert
}\right) \\
&&\times G_{I,X}\left( p_{\Vert },-\tau \right) G_{J,X}\left( p_{\Vert
},\tau \right) ,  \notag
\end{eqnarray}%
where the single-particle Matsubara Green's function is defined by%
\begin{equation}
G_{I,X}\left( p_{\Vert },\tau \right) =-\left\langle T_{\tau }d_{I}\left(
p_{\Vert },\tau \right) d_{I}^{\dag }\left( p_{\Vert },0\right)
\right\rangle .
\end{equation}%
Fourier-transforming $\chi _{\alpha ,\beta }\left( \tau \right) ,$ we get
the familiar result%
\begin{eqnarray}
\chi _{\alpha ,\beta }\left( i\Omega _{p}\right) &=&\int_{0}^{\beta \hslash
}d\tau e^{i\Omega _{p}\tau }\chi _{\alpha ,\beta }\left( \tau \right)
\label{chitau} \\
&=&\frac{e^{2}v_{F}^{2}}{\beta \hslash ^{2}V}\sum_{\omega
_{n}}\sum_{I,J}\sum_{X,p_{\Vert }}\Upsilon _{I,J}^{\left( \alpha \right)
}\left( p_{\Vert }\right) \Upsilon _{J,I}^{\left( \beta \right) }\left(
p_{\Vert }\right)  \notag \\
&&\times G_{I,X}\left( p_{\Vert },i\omega _{n}\right) G_{J,X}\left( p_{\Vert
},i\Omega _{p}+i\omega _{n}\right) ,  \notag
\end{eqnarray}%
where $\Omega _{p},\omega _{n}$ are respectively bosonic and fermionic
Matsubara frequencies. The current-current response has contributions from
both intra- and inter-Landau level transitions. We compute them separately
in the following sections. Moreover, the total current response and the
related dielectric tensor are obtained by summing the individual current
response of the four nodes.

\subsection{Inter-Landau-level contributions to the current response
function $\protect\chi _{\protect\alpha ,\protect\beta }\left( \protect%
\omega \right) $}

With the Hamiltonian given by Eq. (\ref{hamiltone}), the single-particle
Matsubara Green's function is simply 
\begin{equation}
G_{I,X}\left( p_{\Vert },i\omega _{n}\right) =\frac{1}{i\omega _{n}-\left(
E_{I}\left( p_{\Vert }\right) -\mu \right) /\hslash },
\end{equation}%
where $\mu $ is the chemical potential. It is independent of the
guiding-center index $X.$

Performing the frequency sum in Eq. (\ref{chitau}) and taking the analytic
continuation $i\Omega _{p}\rightarrow \omega +i\delta ,$ we get for each
node $\tau $ the interband response function 
\begin{eqnarray}
\chi _{\alpha ,\beta }^{R}\left( \omega \right) &=&-\frac{e^{2}v_{F}^{2}}{%
4\pi ^{2}\ell ^{2}\hslash ^{2}}\sum_{\substack{ I,J  \\ \left( I\neq
J\right) }}\int dp_{\Vert }\Upsilon _{I,J}^{\left( \alpha \right) }\left(
p_{\Vert }\right) \Upsilon _{J,I}^{\left( \beta \right) }\left( p_{\Vert
}\right)  \notag \\
&&\times \frac{f\left( E_{J}\left( p_{\Vert }\right) \right) -f\left(
E_{I}\left( p_{\Vert }\right) \right) }{\omega +i\delta -\left( E_{J}\left(
p_{\Vert }\right) -E_{I}\left( p_{\Vert }\right) \right) /\hslash },
\end{eqnarray}%
where $f\left( E\right) =1/(e^{\beta \left( E-\mu \right) }+1)$ is the Fermi
function, $\beta =1/k_{B}T$ and $k_{B}$ is the Boltzmann constant. At $T=0$
K, $\mu \rightarrow E_{F}$ and $f\left( E_{J}\left( p_{\Vert }\right)
\right) \rightarrow \Theta \left( E_{F}-E_{J}\left( p_{\Vert }\right)
\right) ,$ where $\Theta \left( x\right) $ is the step function and $E_{F}$
the Fermi level. Since disorder is not included in our analysis, each Landau
level is either filled or empty. It follows that only transitions between an
occupied and an unoccupied level can contribute to the response function.

\subsection{Intra-Landau-level contribution to the response function $%
\protect\chi _{\protect\alpha ,\protect\beta }\left( \protect\omega \right) $%
}

We consider the situation where the Fermi level lies in the chiral level $%
n=0,$ above $E=0$ but below the energy of the level $n=1$. The only allowed
intra-Landau level transitions are then those that take place in the chiral
level. To calculate their contribution to the current response function,
disorder needs to be considered. The spectral representation of the
disorder-averaged single-particle Green's function in level $n=0$ is given by%
\begin{equation}
\left\langle G_{0,X}\left( p_{\Vert },i\omega _{n}\right) \right\rangle
=\int_{-\infty }^{+\infty }d\omega \frac{A_{0}\left( p_{\Vert },\omega
\right) }{i\omega _{n}-\omega },
\end{equation}%
with the spectral weight approximated by the Lorentzian shape 
\begin{equation}
A_{0}\left( p_{\Vert },\omega \right) =\frac{\Gamma /\pi }{\left( \omega
-\left( E_{0}\left( p_{\Vert }\right) -\mu \right) /\hslash \right)
^{2}+\Gamma ^{2}},
\end{equation}%
where $\Gamma =1/2\tau _{0},$ with $\tau _{0}$ the momentum relaxation time.
The current response function becomes%
\begin{eqnarray}
\chi _{\alpha ,\beta }\left( i\Omega _{p}\right) &=&\frac{%
e^{2}v_{F}^{2}N_{\varphi }}{\hslash V}\sum_{p_{\Vert }}\Upsilon
_{0,0}^{\left( \alpha \right) }\left( p_{\Vert }\right) \Upsilon
_{0,0}^{\left( \beta \right) }\left( p_{\Vert }\right) \\
&&\times \frac{1}{\beta \hslash }\sum_{\omega _{n}}\int_{-\infty }^{+\infty
}d\omega ^{\prime }\frac{A_{0}\left( p_{\parallel },\omega ^{\prime }\right) 
}{i\Omega _{p}+i\omega _{n}-\omega ^{\prime }}  \notag \\
&&\times \int_{-\infty }^{+\infty }d\omega ^{\prime \prime }\frac{%
A_{0}\left( p_{\parallel },\omega ^{\prime \prime }\right) }{i\omega
_{n}-\omega ^{\prime \prime }}.  \notag
\end{eqnarray}%
Performing the Matsubara frequency sum and taking the analytical
continuation $i\Omega _{p}\rightarrow \omega +i\delta $, we get%
\begin{eqnarray}
\chi _{\alpha ,\beta }^{R}\left( \omega \right) &=&\frac{e^{2}v_{F}^{2}}{%
2\pi \ell ^{2}\hslash ^{2}}\int \frac{dp_{\Vert }}{2\pi }\Upsilon
_{0,0}^{\left( \alpha \right) }\left( p_{\Vert }\right) \Upsilon
_{0,0}^{\left( \beta \right) }\left( p_{\Vert }\right) \\
&&\times \int_{-\infty }^{+\infty }d\omega ^{\prime }\int_{-\infty
}^{+\infty }d\omega ^{\prime \prime }a_{0}\left( p_{\Vert },\omega ^{\prime
}\right)  \notag \\
&&\times a_{0}\left( p_{\parallel },\omega ^{\prime \prime }\right) \frac{%
f\left( \omega ^{\prime \prime }\right) -f\left( \omega ^{\prime }\right) }{%
\omega +i\delta +\omega ^{\prime \prime }-\omega ^{\prime }},  \notag
\end{eqnarray}%
where we have defined 
\begin{equation}
a_{0}\left( p_{\Vert },\omega \right) =A_{0}\left( p_{\Vert },\omega +\mu
/\hslash \right) .
\end{equation}%
Defining also%
\begin{equation}
g_{0}\left( p_{\Vert },\omega \right) =\frac{\omega -E_{0}\left( p_{\Vert
}\right) /\hslash }{\left( \omega -E_{0}\left( p_{\Vert }\right) /\hslash
\right) ^{2}+\Gamma ^{2}},
\end{equation}%
we get, at zero temperature and after some simple algebra

\begin{eqnarray}
\chi _{\alpha \beta }^{R}\left( \omega \right) &=&\frac{e^{2}v_{F}^{2}}{2\pi
\ell ^{2}\hslash ^{2}}\int \frac{dp_{\Vert }}{2\pi }\Upsilon _{0,0}^{\left(
\alpha \right) }\left( p_{\Vert }\right) \Upsilon _{0,0}^{\left( \beta
\right) }\left( p_{\Vert }\right) \\
&&\times \left[ \int_{-\infty }^{+E_{F}/\hslash }d\omega ^{\prime
}a_{0}\left( p_{\Vert },\omega ^{\prime }\right) \right.  \notag \\
&&\times \left[ g_{0}\left( p_{\Vert },\omega ^{\prime }+\omega \right)
+g_{0}\left( p_{\Vert },\omega ^{\prime }-\omega \right) \right]  \notag \\
&&\left. -i\pi \int_{E_{F}/\hslash -\omega }^{E_{F}/\hslash }d\omega
^{\prime }a_{0}\left( p_{\Vert },\omega ^{\prime }\right) a_{0}\left(
p_{\Vert },\omega ^{\prime }+\omega \right) \right] .  \notag
\end{eqnarray}%
Thus, the intraband conductivity in the chiral level is given by

\begin{eqnarray}
\sigma _{\alpha ,\beta }^{intra}\left( \omega \right) &=&\frac{e^{2}v_{F}^{2}%
}{2\pi \ell ^{2}\hslash ^{2}}\frac{i}{\omega }\int \frac{dp_{\Vert }}{2\pi }%
\Upsilon _{0,0}^{\left( \alpha \right) }\left( p_{\Vert }\right) \Upsilon
_{0,0}^{\left( \beta \right) }\left( p_{\Vert }\right)  \label{sigmaintra} \\
&&\times \left[ \int_{-\infty }^{+E_{F}/\hslash }d\omega ^{\prime
}a_{0}\left( p_{\Vert },\omega ^{\prime }\right) K\left( p_{\Vert },\omega
,\omega ^{\prime }\right) \right.  \notag \\
&&\left. -i\pi \int_{E_{F}/\hslash -\omega }^{E_{F}/\hslash }d\omega
^{\prime }a_{0}\left( p_{\Vert },\omega ^{\prime }\right) a_{0}\left(
p_{\Vert },\omega ^{\prime }+\omega \right) \right] .  \notag
\end{eqnarray}%
with the function%
\begin{eqnarray}
K\left( p_{\Vert },\omega ,\omega ^{\prime }\right) &=&g_{0}\left( p_{\Vert
},\omega ^{\prime }+\omega \right) +g_{0}\left( p_{\Vert },\omega ^{\prime
}-\omega \right) \\
&&-2\delta _{\alpha ,\beta }g_{0}\left( p_{\Vert },\omega ^{\prime }\right) .
\notag
\end{eqnarray}

In the absence of a tilt, only the matrix element $\Upsilon _{0,0}^{\left(
\alpha \right) }\left( p_{\Vert }\right) $ for $\alpha $ in the direction of
the magnetic field is nonzero. Thus, only the conductivity $\sigma _{\Vert
}\left( \omega \right) $ (which is $\sigma _{xx}\left( \omega \right) $ for $%
\mathbf{B}=B_{0}\widehat{\mathbf{x}}$ and $\sigma _{zz}\left( \omega \right) 
$ for $\mathbf{B}=B_{0}\widehat{\mathbf{z}}$) is nonzero. From Eq. (\ref%
{sigmaintra}), we get 
\begin{eqnarray}
\func{Re}\left[ \sigma _{\Vert }\left( \omega \right) \right] &=&\frac{%
v_{F}e^{3}\tau _{0}}{4\pi ^{2}\hslash ^{2}}\frac{B_{0}}{1+\left( \omega \tau
_{0}\right) ^{2}},  \label{sigiii} \\
\func{Im}\left[ \sigma _{\Vert }\left( \omega \right) \right] &=&\omega \tau
_{0}\func{Re}\left[ \sigma _{\Vert }\left( \omega \right) \right] .
\end{eqnarray}%
Our results for the conductivity contains the momentum instead of the
transport relaxation time since vertex corrections are not included in our
calculation of the current response function.

A finite tilt modifies the matrix elements $\Upsilon _{0,0}^{\left( \alpha
\right) }\left( p_{\Vert }\right) $ and makes the other elements of the
conductivity tensor nonzero, in particular $\sigma _{\bot }\left( \omega
\right) $ (i.e. $\sigma _{zz}\left( \omega \right) $ for $\mathbf{B}=B_{0}%
\widehat{\mathbf{x}}$ and $\sigma _{xx}\left( \omega \right) $ for $\mathbf{B%
}=B_{0}\widehat{\mathbf{z}}$), which enters in the definition of the Kerr
angle for the Voigt configuration, as we show below.

With $\tau _{0}$ and $v_{F}$ independent of the magnetic field, Eq. (\ref%
{sigiii}) shows that the conductivity $\func{Re}\left[ \sigma _{\Vert
}\left( \omega \right) \right] $ increases linearly with the magnetic field.
This negative magnetoresistance is a signature of the chiral anomaly in Weyl
semimetals\cite{Chiral anomaly}, where collinear electric and magnetic
fields ($\mathbf{E}\cdot \mathbf{B}\neq 0$) result in a transport of
electrons between two Weyl nodes of opposite chirality. It is known,
however, that $\tau _{0}$ varies with magnetic field in a way that depends
on the type of disorder considered. For instance, in the case of
short-ranged neutral impurities, $\func{Re}\left[ \sigma _{\Vert }\left(
\omega \right) \right] $ becomes independent of $B.$ The $B$-dependence of $%
v_{F}$ can further alter the sign of the magnetoresistance\cite%
{DasSarma1,Lou2015}. As a result, the $B$-dependence of $\func{Re}\left[
\sigma _{\Vert }\left( \omega \right) \right] $ is not robustly linked to
the chiral anomaly.

\begin{figure}
\centering\includegraphics[width = \linewidth]{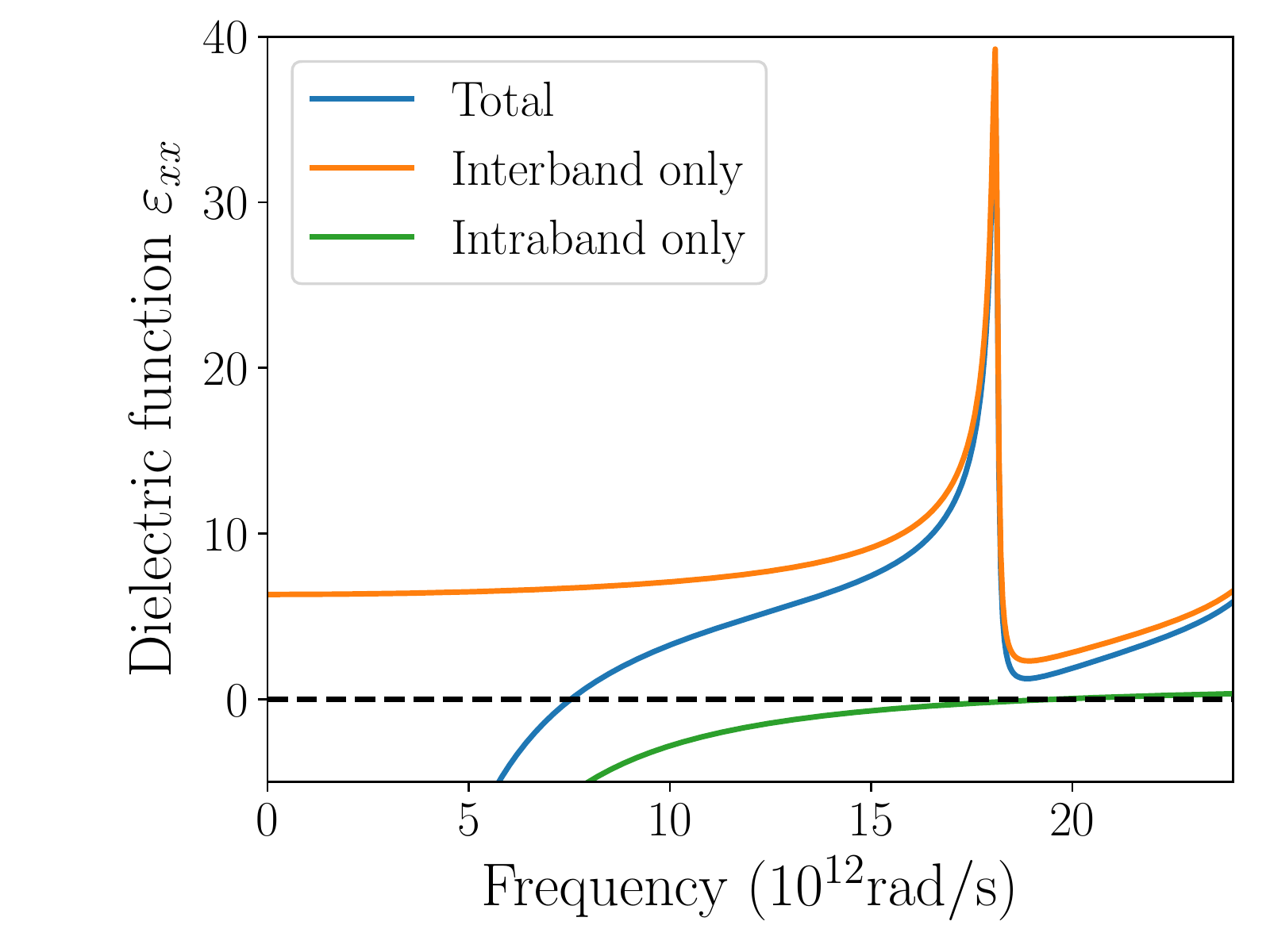} 
\caption{Real part of the relative dielectric
		function $\protect\varepsilon _{\Vert } =\protect\varepsilon _{xx}  $ as a
		function of $\protect\omega $ for an external static magnetic field
		$\mathbf{B}=B_{0}\protect\widehat{\mathbf{x}}$ with $B_{0}=0.3$T$.$ 
		The zero of $\protect\varepsilon _{\Vert }\left(\protect\omega \right)$, which gives the plasmon frequency, is pushed to a
		lower frequency when inter-Landau level contributions are considered. 
		Parameters: total electronic density $n_{e}=1\times 10^{20}$ m$^{-3}$, momentum relaxation time
		$\protect\tau _{0}=10$ ps, Fermi velocity
		$v_{F}=3\times 10^{5}$ m/s, and tilt $t=0.$} \label{fig4}
\end{figure}

In contrast, the situation appears to be more promising when it comes to the
plasmon frequency $\omega _{p}$. At $q=0$, $\omega_p$ is given by the
condition 
\begin{equation}
\func{Re}\left[ \varepsilon _{||}\left( \omega _{p}\right) \right] =0,
\end{equation}%
i.e. 
\begin{equation}
1-\frac{\func{Im}\left[ \sigma _{\Vert }\left( \omega _{p}\right) \right] }{%
\varepsilon _{0}\omega _{p}}=0,
\end{equation}%
which gives%
\begin{equation}
1-\frac{1}{\varepsilon _{0}}\frac{v_{F}e^{3}}{4\pi ^{2}\hslash ^{2}}\frac{%
\tau _{0}^{2}}{1+\left( \omega _{p}\tau _{0}\right) ^{2}}B_{0}=0.
\end{equation}%
The plasmon frequency occurs at high frequency (in the THz range in WSMs),
which can exceed the momentum scattering rate. When $\omega _{p}\tau _{0}\gg
1$, one has 
\begin{equation}
\omega _{p}^{2}=\frac{e^{2}v_{F}}{4\pi ^{2}\hslash \ell ^{2}\varepsilon _{0}}%
,  \label{plasmon}
\end{equation}%
where $\varepsilon _{0}$ is the permittivity of free space (in general to be
multiplied by $\varepsilon _{\infty }$ due to screening from high-energy
electronic bands). The linear increase with magnetic field of $\omega
_{p}^{2}$ is also a signature of the chiral anomaly\cite{Chiral anomaly} and
should not be modified by disorder insofar as $\omega _{p}\tau _{0}\gg 1$.
This trend remains likewise robust to the $B-$dependence of $v_F$ (neglected
herein). In a normal metal, the plasmon frequency $\omega
_{p,metal}^{2}=n_{e}e^{2}/m\varepsilon _{0},$ with $n_{e}$ the electronic
density and $m$ the effective mass of the electron, is independent of the
magnetic field (see Appendix C). The dispersion relation of the plasmon mode
in WSMs and normal metals are discussed in more detail in Ref. %
\onlinecite{plasmon2}. The plasmon frequency, or equivalently the zero of
the longitudinal dielectric function, increases as $\sqrt{B_{0}}$. When
interband transitions are considered, $\omega _{p}$ is shifted to a lower
frequency as shown in Fig. \ref{fig4}, although it remains in the THz range.

\section{FORMALISM\ FOR\ THE\ MAGNETO-OPTICAL KERR EFFECT}

We consider an electromagnetic wave arriving at normal incidence on a
surface of a semi-infinite WSM that has no Fermi arcs. The Maxwell equations
in region 1 (the vacuum) and 2 (the WSM) are 
\begin{eqnarray}
\mathbf{\nabla \cdot D} &=&\rho _{f};\ \mathbf{\nabla \times E}=-\frac{%
\partial \mathbf{B}}{\partial t}, \\
\ \mathbf{\nabla \cdot B} &=&0;\mathbf{\nabla \times H}=\mathbf{j}_{f}+\frac{%
\partial \mathbf{D}}{\partial t}\mathbf{,}
\end{eqnarray}%
where $\rho _{f}$ and $\mathbf{j}_{f}$ are the free charge and current
densities. We assume that the WSM\ is non-magnetic so that $\mathbf{B}=\mu
_{0}\mathbf{H}$ with $\mu _{0}$ the permeability of free space. We use the
constitutive relation $\mathbf{D}=\epsilon _{0}\overleftrightarrow{\mathbf{%
\varepsilon }}\cdot \mathbf{E,}$ where $\overleftrightarrow{\mathbf{%
\varepsilon }}$ is the \textit{relative} dielectric tensor which is related
to the conductivity tensor $\overleftrightarrow{\mathbf{\sigma }}$ by 
\begin{equation}
\overleftrightarrow{\mathbf{\varepsilon }}=\overleftrightarrow{\mathbf{I}}+i%
\frac{\overleftrightarrow{\mathbf{\sigma }}}{\varepsilon _{0}\omega },
\end{equation}%
with $\overleftrightarrow{\mathbf{I}}$ the unit tensor and $\varepsilon _{0}$
the permittivity of free space. To account for the high-energy transitions
not included in our calculation, the unit tensor $\overleftrightarrow{%
\mathbf{I}}$ should be replaced by the relative dielectric tensor $%
\overleftrightarrow{\mathbf{\varepsilon }}_{\infty }$ whose precise form
depends on the particular WSM\ considered. In our calculation, we take $%
\overleftrightarrow{\mathbf{\varepsilon }}_{\infty }=\varepsilon _{\infty }%
\overleftrightarrow{\mathbf{I}}$ with $\varepsilon _{\infty }=1.$ Because $%
\varepsilon _{\infty }$ can be quite big in WSMs, we discuss the effects of
increasing $\varepsilon _{\infty }$ on the Kerr angle in Sec. VI.

From the Maxwell equations, the dispersion relation of an electromagnetic
wave in both regions (with $\overleftrightarrow{\mathbf{\varepsilon }}=%
\overleftrightarrow{\mathbf{I}}$ in region 1) is given by

\begin{equation}
q^{2}\mathbf{E-}\left( \mathbf{q\cdot E}\right) \mathbf{q=}\frac{\omega ^{2}%
}{c^{2}}\overleftrightarrow{\mathbf{\varepsilon }}\cdot \mathbf{E}.
\label{eigen_eq}
\end{equation}

At the interface between vacuum and WSM, the electromagnetic field must obey
the boundary conditions%
\begin{eqnarray}
\left( \mathbf{D}_{1}\mathbf{-D}_{2}\right) \cdot \mathbf{\hat{n}} &=&\sigma
_{f},  \label{boundary_conditions} \\
\left( \mathbf{B}_{1}\mathbf{-B}_{2}\right) \cdot \mathbf{\hat{n}} &=&0, 
\notag \\
\mathbf{E}_{1}^{\parallel }\mathbf{-E}_{2}^{\parallel } &=&\mathbf{0}, 
\notag \\
\mathbf{H}_{1}^{\parallel }\mathbf{-H}_{2}^{\parallel } &=&\mathbf{0}, 
\notag
\end{eqnarray}%
where $\mathbf{\hat{n}}$ is a vector normal to the WSM surface, pointing
from region 2 to 1. $\mathbf{E}_{i}^{\parallel },\mathbf{H}_{i}^{\parallel }$
(with $i=1,2$) are the field components parallel to the surface of the WSM,
which we take to be the $x-y$ plane at $z=0$.

The incident wave is assumed to be linearly polarized in the $x-y$ plane :\ $%
\mathbf{E}_{I}(z,t)=(E_{0}^{x}\mathbf{\hat{x}}+E_{0}^{y}\mathbf{\hat{y}}%
)e^{i(qz-\omega t)}.$ Its dispersion relation is $q^{2}=\omega ^{2}/c^{2}$,
where $c$ is the speed of light in vacuum. The reflected wave is given by 
\begin{align}
\mathbf{E}_{R}(z,t)=& [(r_{xx}E_{0}^{x}+r_{xy}E_{0}^{y})\mathbf{\hat{x}} \\
& +(r_{yx}E_{0}^{x}+r_{yy}E_{0}^{y})\mathbf{\hat{y}}]e^{-i(qz+\omega t)}, 
\notag
\end{align}%
where $r_{ij}$ are complex reflection coefficients that depend on the
orientation of the magnetic field, tilt vector and mirror plane. We consider
two different cases in the following sections: the longitudinal and
transverse configurations.

\subsection{Longitudinal propagation $\left( \mathbf{q\parallel
B_{0}\parallel \hat{z}}\right) $}

In the longitudinal configuration, the electromagnetic wave propagates in
the direction of the external magnetic field. If the external magnetic field 
$\mathbf{B_{0}}$ and the tilt vector $\mathbf{t}$ are in the $\mathbf{\hat{z}%
}$ direction, perpendicular to the mirror plane, then the total dielectric
tensor (sum of the four nodes) has the form%
\begin{equation}
\overleftrightarrow{\mathbf{\varepsilon }}=\left( 
\begin{array}{ccc}
\varepsilon _{xx} & \varepsilon _{xy} & 0 \\ 
-\varepsilon _{xy} & \varepsilon _{xx} & 0 \\ 
0 & 0 & \varepsilon _{zz}%
\end{array}%
\right) .  \label{tensor}
\end{equation}%
In this situation, Maxwell equations support in the WSM\ two elliptically
polarized electromagnetic waves, the analog of the right (RCP) and left
(LCP)\ circularly polarized waves, with electric fields $E_{+}=E_{x}+iE_{y}$
and $E_{-}=E_{x}-iE_{y}$. The transmitted electric fields for these two
solutions are given by : 
\begin{align}
\mathbf{E}_{T}^{\left( \pm \right) }(z,t)=& [(t_{xx}^{\left( \pm \right)
}E_{0}^{x}+t_{xy}^{\left( \pm \right) }E_{0}^{y})\mathbf{\hat{x}} \\
& +(t_{yx}^{\left( \pm \right) }E_{0}^{x}+t_{yy}^{\left( \pm \right)
}E_{0}^{y})\mathbf{\hat{y}}]e^{-i(q_{\pm }z-\omega t)},  \notag
\end{align}%
where $t_{ij}(\omega )$ are complex transmission coefficients. The component 
$E_{T,z}=0$ so that the wave is transverse in both media and there is no
induced charge (i.e. $\mathbf{\nabla \cdot D}=0$) in the WSM. The dispersion
relations are

\begin{eqnarray}
q_{\pm }^{2}(\omega ) &=&\frac{\omega ^{2}}{c^{2}}\left( \varepsilon
_{xx}\pm i\varepsilon _{xy}\right)  \label{eq:q2_pm} \\
&\equiv &\frac{\omega ^{2}}{c^{2}}\varepsilon _{\mp }\left( \omega \right) 
\notag \\
&=&q^{2}\varepsilon _{\mp }\left( \omega \right)  \notag
\end{eqnarray}%
with $q=\omega /c.$

The total electric fields in regions 1 and 2 are given by%
\begin{eqnarray}
\mathbf{E}_{1}(z,t) &=&\mathbf{E}_{I}(z,t)+\mathbf{E}_{R}(z,t), \\
\mathbf{E}_{2}(z,t) &=&\mathbf{E}_{T}^{+}(z,t)+\mathbf{E}_{T}^{-}(z,t),
\end{eqnarray}%
with the magnetic fields obtained from Faraday's law$.$

Applying the boundary conditions to the total electric and magnetic field
and isolating the reflection coefficients, we get 
\begin{subequations}
\label{eq:reflection_matrix}
\begin{align}
r_{xx}(\omega )& =r_{yy}(\omega )=\frac{1-\sqrt{\varepsilon _{+}\varepsilon
_{-}}}{\left( 1+\sqrt{\varepsilon _{+}}\right) \left( 1+\sqrt{\varepsilon
_{-}}\right) },  \label{rxx} \\
r_{yx}(\omega )& =-r_{xy}(\omega )=\frac{i\left( \sqrt{\varepsilon _{+}}-%
\sqrt{\varepsilon _{-}}\right) }{\left( 1+\sqrt{\varepsilon _{+}}\right)
\left( 1+\sqrt{\varepsilon _{-}}\right) }.  \label{rxy}
\end{align}
\end{subequations}

In this configuration, the incident wave is taken to be linearly polarized
at an angle $\theta _{I}=\pi /4$ from the $x$ axis with amplitude $E_{0}^{x}$
along both directions of the $x-y$ plane, so that the reflected electric
field is 

\begin{equation}
\mathbf{E}_{R}(z,t)=\left( E_{R}^{x}(\omega )\mathbf{\hat{x}}%
+E_{R}^{y}(\omega )\widehat{\mathbf{y}}\right) e^{-i(qz+\omega t)},
\end{equation}%
with%
\begin{eqnarray}
E_{R}^{x}(\omega ) &=&\left(r_{xx}(\omega ) + r_{xy}(\omega
)\right)E_{0}^{x}, \\
E_{R}^{y}(\omega ) &=&\left(r_{xx}(\omega ) - r_{xy}(\omega
)\right)E_{0}^{x}.
\end{eqnarray}%
The coefficients $r_{ij}(\omega )$ are obtained from the dielectric tensor $%
\overleftrightarrow{\mathbf{\varepsilon }},$ which contains the combined
current response function of the four nodes.

\subsection{Transverse propagation $\left( \mathbf{q\perp \mathbf{B_{0}}%
\parallel \hat{x}}\right) $}

In the transverse configuration, the electromagnetic wave still propagates
along the $z$ axis, but the external magnetic field is taken to point in the
direction $\mathbf{\hat{x},}$ i.e. parallel to the mirror plane. For a tilt
along the $z$ axis, the dielectric tensor for this configuration has the form

\begin{equation}
\overleftrightarrow{\varepsilon }=\left( 
\begin{array}{ccc}
\varepsilon _{xx} & 0 & 0 \\ 
0 & \varepsilon _{yy} & \varepsilon _{yz} \\ 
0 & -\varepsilon _{yz} & \varepsilon _{zz}%
\end{array}%
\right)  \label{tensor2}
\end{equation}%
when the contributions of the four nodes are taken into account.

Maxwell's equations give again two dispersion relations: one where the
electric field $\mathbf{E}_{\Vert }$ of the incident light is polarized
along $x,$ i.e. parallel to the static magnetic field, giving the dispersion
relation

\begin{equation}
q_{\parallel }^{2}=\frac{\omega ^{2}}{c^{2}}\varepsilon _{xx},
\label{qparperp}
\end{equation}%
and one where the electromagnetic wave is polarized along $y,$ i.e.
perpendicular to the static magnetic field with the dispersion relation

\begin{equation}
q_{\perp }^{2}=\frac{\omega ^{2}}{c^{2}}\left( \varepsilon _{yy}+\frac{%
\varepsilon _{yz}^{2}}{\varepsilon _{zz}}\right) =\frac{\omega ^{2}}{c^{2}}%
\varepsilon _{v},  \label{qparperpo}
\end{equation}%
where $\varepsilon _{v}$ is called the Voigt dielectric function\cite%
{Balkansky}. In this later case, there is an induced field in the direction
of propagation, given by 
\begin{equation}
E_{z}=-\frac{\varepsilon _{zy}}{\varepsilon _{zz}}E_{y},
\end{equation}%
so that the electric field for this polarization, in medium 2, is given by%
\begin{equation}
\mathbf{E}_{\bot }=E_{y}\left( \widehat{\mathbf{y}}-\frac{\varepsilon _{zy}}{%
\varepsilon _{zz}}\widehat{\mathbf{z}}\right) .
\end{equation}%
The induced charge is still zero, however. The transmitted wave in the WSM
is given by%
\begin{equation}
\mathbf{E}_{T}^{\left( \pm \right) }(z,t)=t_{\Vert }\mathbf{E}_{\Vert
}e^{-i(q_{\Vert }z-\omega t)}+t_{\bot }\mathbf{E}_{\bot }e^{-i(q_{\bot
}z-\omega t)},
\end{equation}%
where $t_{\Vert },t_{\bot }$ are the transmission coefficients for the
parallel and transverse polarizations. In the Voigt configuration, the
polarization vector of an incident electromagnetic wave polarized at an
angle with respect to the $x$ axis in the $x-y$ plane will rotate in the WSM
because of the different refractive indices for the parallel and transverse
polarizations. There is no rotation if the polarization vector is only along
the $x$ or $y$ axis. This form of dichroism can also lead to a sizeable Kerr
rotation as we show in Sec. V. Solving Eqs. (\ref{eigen_eq}) and (\ref%
{boundary_conditions}) gives the reflected field components

\begin{eqnarray}
E_{R}^{x}(\omega ) &=&\frac{1-\sqrt{\varepsilon _{xx}}}{1+\sqrt{\varepsilon
_{xx}}}E_{0}^{x}(\omega ),  \label{fields} \\
E_{R}^{y}(\omega ) &=&\frac{1-\sqrt{\varepsilon _{v}}}{1+\sqrt{\varepsilon
_{v}}}E_{0}^{y}(\omega ).  \notag
\end{eqnarray}

\subsection{Definition of the Kerr angle}

We define the phases $\theta _{R}^{x}$ and $\theta _{R}^{y}$ by 
\begin{equation}
E_{R}^{x}=\left\vert E_{R}^{x}\right\vert ^{i\theta
_{R}^{x}},\;E_{R}^{y}=\left\vert E_{R}^{y}\right\vert e^{i\theta _{R}^{y}},
\label{phases}
\end{equation}%
where $\left\vert E_{R}^{i}\right\vert ,\theta _{R}^{i}$ (with $i=x,y$) are
respectively the moduli and associated phases of the complex components of
the reflected wave. The reflected electric field at $z=0$ is then%
\begin{equation}
\mathbf{E}\left( t\right) =\left\vert E_{R}^{x}\right\vert \cos \left(
\theta _{R}^{x}-\omega t\right) \widehat{\mathbf{x}}+\left\vert
E_{R}^{y}\right\vert \cos \left( \theta _{R}^{y}-\omega t\right) \widehat{%
\mathbf{y}}.
\end{equation}%
The polarization of the reflected field is strongly modified as the
frequency is varied. It could be linear, elliptical or circular. The sense
of rotation of the polarization can also change with $\omega .$ A complete
description of the reflected field would give $\mathbf{E}\left( t\right) $
over one period of oscillation. Experimentally, however, what is reported is
the Kerr angle $\theta _{K}$ and ellipticity $\phi _{K}$ defined by 
\begin{equation}
\tan \left( \theta _{K}+\theta _{I}\right) =\left\vert \frac{E_{R}^{y}}{%
E_{R}^{x}}\right\vert ,  \label{Standard}
\end{equation}%
where $\theta _{I}$ is the polarization angle of the incident linearly
polarized wave, and the ellipticity%
\begin{equation}
\phi _{K}=\theta _{R}^{x}-\theta _{R}^{y}.
\end{equation}

In the Faraday configuration, with an incident wave linearly polarized at an
angle $\theta _{I}=\pi /4$ with respect to the $x$ axis, the Kerr angle is
given by

\begin{equation}
\tan \left( \theta _{K}+\theta _{I}\right) =\left\vert \frac{1+i\left( \sqrt{%
\varepsilon _{+}}-\sqrt{\varepsilon _{-}}\right) -\sqrt{\varepsilon
_{+}\varepsilon _{-}}}{1-i\left( \sqrt{\varepsilon _{+}}-\sqrt{\varepsilon
_{-}}\right) -\sqrt{\varepsilon _{+}\varepsilon _{-}}}\right\vert ,
\label{tanfara}
\end{equation}%
while in the Voigt configuration for the identically polarized incident
wave, the Kerr angle is 
\begin{equation}
\tan \left( \theta _{K}+\theta _{I}\right) =\left\vert \frac{\left( 1-\sqrt{%
\varepsilon _{v}}\right) \left( 1+\sqrt{\varepsilon _{xx}}\right) }{\left( 1+%
\sqrt{\varepsilon _{v}}\right) \left( 1-\sqrt{\varepsilon _{xx}}\right) }%
\right\vert .  \label{tanvoigt}
\end{equation}

\section{NUMERICAL RESULTS FOR THE KERR ANGLE}

We compute the Kerr angle in the four-node model for two specific
configurations: (1) the longitudinal Faraday configuration where $\mathbf{%
q\parallel B}_{0}\mathbf{\parallel \hat{z}}$ and the linear polarization
vector is $\widehat{\mathbf{e}}\bot \mathbf{B}_{0};$ and (2)\ the transverse
Voigt configuration where $\mathbf{q\parallel \hat{z},}$ $\mathbf{B}%
_{0}=B_{0}\widehat{\mathbf{x}}\bot \mathbf{q}$ and the linear polarization
vector $\widehat{\mathbf{e}}$ is in the $x-y$ plane. For the Faraday
configuration, we choose the magnetic field to be perpendicular to the
mirror plane while we take it in the mirror plane in the Voigt
configuration. Other choices are, of course, possible. We assume a total
electronic density of $n_{e}=1\times 10^{20}$ m$^{-3}$ for the four nodes.
For the interband transitions, we introduce the scattering rate
phenomenologically by taking a finite but small value of $\delta $ in the
conductivity tensor. For the intraband transitions, disorder is taken into
account explicitly by introducing a lorentzian broadening $\Gamma =1/2\tau
_{0}$ with $\tau _{0}=10^{-11}$ s. In both configurations, the tilt vector
is taken to be $\mathbf{t}=t_{0}\widehat{\mathbf{z}}$ with $t_{0}=0.5$, so
that $\mathbf{t}\Vert \mathbf{B}_{0}$ in the Faraday configuration and $%
\mathbf{t}\bot \mathbf{B}_{0}$ in the Voigt configuration. This is done in
order to isolate the main features of the Kerr angle in both configurations,
i.e. valley polarization in the former, and chiral anomaly in the latter. In
the presentation of our results, we change the notation for the Landau level
index, i.e. we choose $n$ to be positive for $s=+1$ and negative for $s=-1.$
The chiral Landau level is still $n=0.$

\subsection{Faraday configuration}

When the external magnetic field has a non-zero projection along the tilt
vector, an optical valley polarization effect appears in the absorption
spectrum. This effect results in a splitting of the interband transition
peaks $0\rightarrow \left\vert n\right\vert $ and $-\left\vert n\right\vert
\rightarrow 0$ that involve the chiral Landau level. One pair of nodes
starts absorbing light before the other pair, leading to a valley
polarization. This phenomenon was studied extensively in our previous work
on the absorption spectrum of a WSM\cite{Bertrand2019}. Here, we show that
it is also possible to probe it using the MOKE.

\begin{figure}
\centering\includegraphics[width = \linewidth]{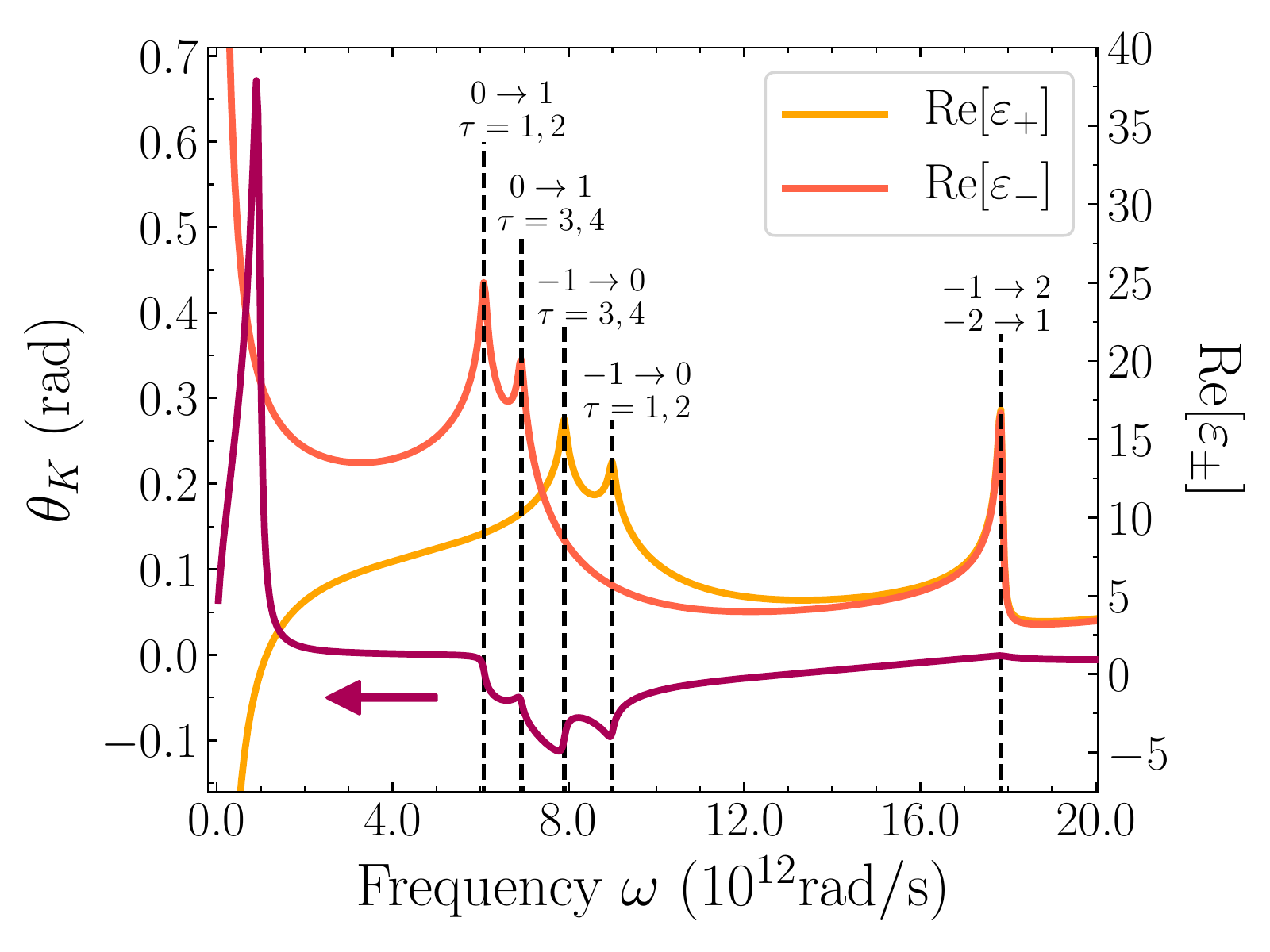} 
\caption{Frequency dependence of the
		Kerr angle $\protect\theta _{K}$ and real part of the dielectric functions $		\protect\varepsilon _{+}$ and $\protect\varepsilon _{-}$ in the Faraday
		configuration, where $\mathbf{q},\mathbf{B}_{0},\mathbf{t\parallel }\protect		\widehat{\mathbf{z}}$. The interband transitions contributing to the main
		peaks are indicated. The $0\rightarrow 1$ and $-1\rightarrow 0$ transitions
		are each split into two as a consequence of valley polarization; the Weyl
		nodes $\protect\tau $ contributing to each of the subpeaks are indicated.
		Parameters: $B_{0}=0.2$T, $n_{e}=1\times 10^{20}$ m$^{-3}$, $		v_{F}=3\times 10^{5}$ m/s, $\mathbf{t}=0.5\protect\widehat{\mathbf{z}}.$}
\label{fig5}
\end{figure}

\begin{figure}
\centering
\includegraphics[width = \linewidth]{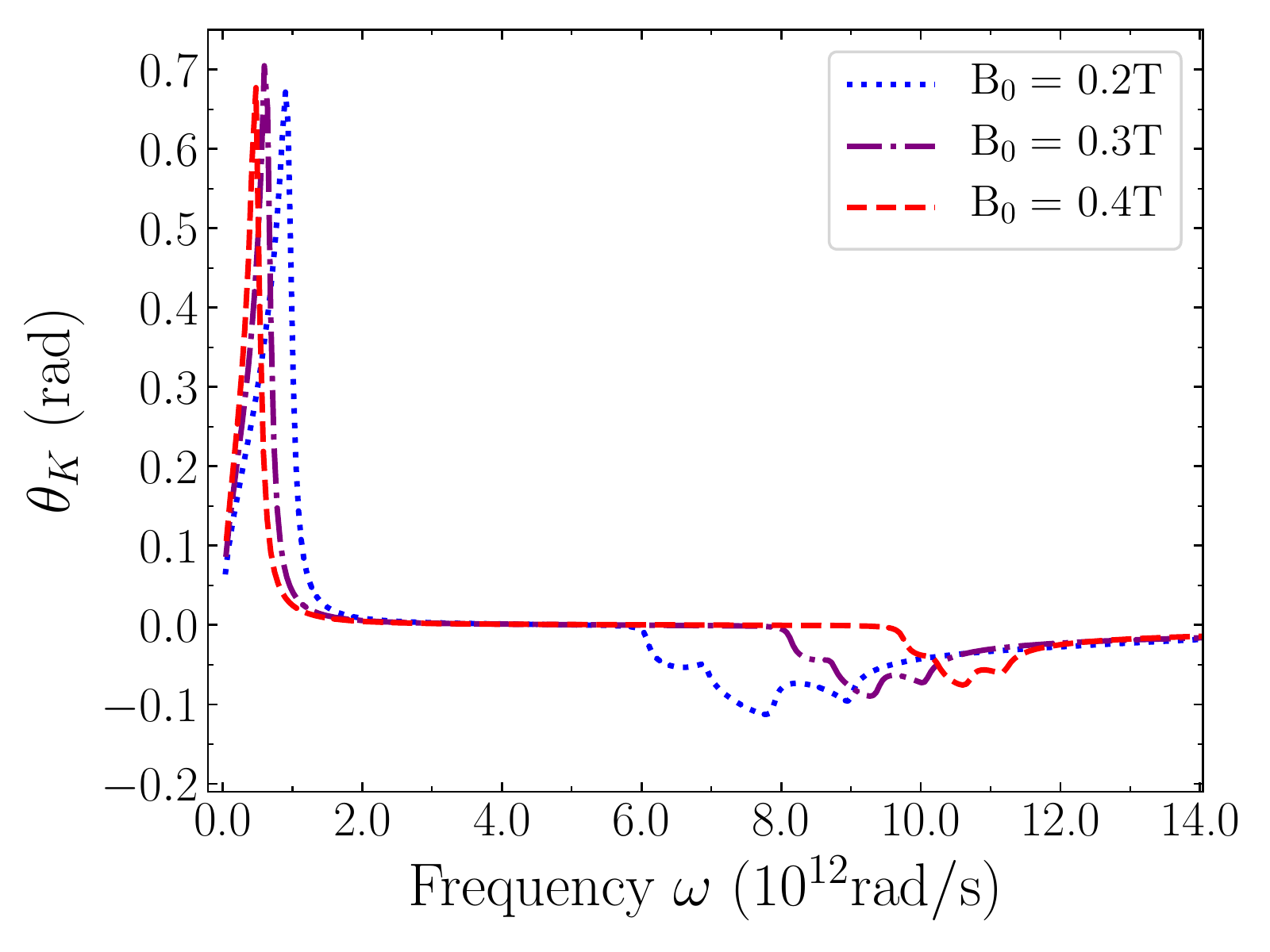} 
\caption{Kerr angle as a function of
		frequency for different values of the magnetic field $\mathbf{B}=B_{0}\protect\widehat{\mathbf{z}}$
		in the Faraday configuration. Parameters: $n_{e}=1\times 10^{20}$ m$^{-3}$, $v_{F}=3\times 10^{5}$ m/s, $\mathbf{t}		=0.5\protect\widehat{\mathbf{z}}.$}
\label{fig6}
\end{figure}

A plot of the Kerr angle versus frequency is shown in Fig. \ref{fig5} for
the case where the vectors $\mathbf{q},\mathbf{B}_{0},\mathbf{t}$ are all
parallel to the $z$ axis and so perpendicular to the mirror plane. The
incident wave is assumed linearly polarized in the $x-y$ plane with $\theta
_{I}=\pi /4$ and the Fermi level is in the chiral Landau level. In this
configuration, only the dipolar transitions are permitted. The allowed
interband transitions are visible as a succession of dips in the curve of
the Kerr angle \textit{vs} frequency for $\omega \gtrsim 4\times 10^{12}$
rad/s or peaks in the dielectric functions $\func{Re}\left[ \varepsilon
_{\pm }\left( \omega \right) \right] $. The curve of $\theta _{K}\left(
\omega \right) $ shows a large peak around $\omega \approx 1\times 10^{12}$
rad/s, which coincides with the frequency $\omega _{+}$ where $\func{Re}%
\left[ \varepsilon _{+}\left( \omega \right) \right] =0$.\textbf{\ }Figure %
\ref{fig6} shows that this large peak in $\theta _{K}\left( \omega \right) $
is redshifted in frequency when the magnetic field increases, which is also
the behavior of the frequency $\omega _{+}\left( B\right) .$ A similar shift
happens in a normal metal (see Appendix C). We remark that $\tau _{0}$ does
not enter in the calculation of the Kerr angle, i.e. in Eq. (\ref{tanfara}),
in this configuration.

Figure \ref{fig5} shows that nodes related by time-reversal symmetry (TRS),
i.e. nodes 1 and 3 or 2 and 4, have different optical gaps in the Faraday
configuration. This causes a splitting of the dips in $\theta _{K}\left(
\omega \right) $ or the peaks in $\func{Re}\left[ \varepsilon _{\pm }\left(
\omega \right) \right] $ corresponding to the transitions $0\rightarrow 1$
and $-1\rightarrow 0$. (Note that we could have chosen a different Faraday
configuration leading to no splitting.) The other interband peaks in $\func{%
Re}\left[ \varepsilon _{\pm }\left( \omega \right) \right] $, corresponding
to transitions between nonchiral Landau levels, are not split. By contrast,
when $\mathbf{t\cdot B}_{0}=0$ for all nodes, transitions $0\rightarrow 1$
(or $-1\rightarrow 0$) are equivalent for all four nodes and therefore, only
a single dip or peak is present for each transition $0\rightarrow 1$ or $%
-1\rightarrow 0$. With a linearly polarized incident wave, both $%
0\rightarrow 1$ and $-1\rightarrow 0$ transitions are excited. For a
circularly polarized light, either the $0\rightarrow 1$ or $-1\rightarrow 0$
transition is excited. (More specifically, the transitions $\left\vert
n\right\vert \rightarrow \left\vert n\right\vert +1$ show up in $\varepsilon
_{-}$ and the transitions $\left\vert n\right\vert \rightarrow \left\vert
n\right\vert -1$ in $\varepsilon _{+}$, as in graphene.)\ The valley
polarization effect due to a finite tilt, for transitions involving the
chiral level, are thus detectable in the Kerr rotation spectra.

Using Eq. (\ref{goerg}), the energy gap between two Landau levels $n,m\neq 0$
of a tilted Weyl cone is independent of the node index and increases with
magnetic field. As we just showed, the gap involving the chiral level
behaves differently. If the Fermi level is not at the neutrality point, it
depends on the node index and the splitting of the $0\rightarrow 1$ (or $%
-1\rightarrow 0$) peak decreases with increasing magnetic field as can be
seen in Fig. \ref{fig6}. This behavior is also present in the absorption
spectrum (see Fig. 9 of Ref. \onlinecite{Bertrand2019}). It follows that, in
the quantum limit, the valley polarization is present in the Kerr rotation
angle in a larger range of frequencies as the magnetic field is decreased.
The splitting can also be increased by increasing the tilt. It would be
interesting to measure these effects experimentally.

\subsection{Voigt configuration}

In the Voigt configuration, the magnetic field $\mathbf{B}=B_{0}\widehat{%
\mathbf{x}}$ is perpendicular to the propagation vector $\mathbf{q\parallel 
\hat{z}}$ and the tilt vector $\mathbf{t\parallel \hat{z}}$ and the
polarization vector $\widehat{\mathbf{e}}$ lies in the $x-y$ plane. As Eq. (%
\ref{fields}) shows, in order to see a Kerr rotation in the Voigt
configuration, the polarization of the incident wave propagating along the $%
z $ direction must make an angle with the $x$ and $y$ axis. We choose $%
\theta _{I}=\pi /4$ and $E_{0}^{x}=E_{0}^{y}=E_{0}$ for the incoming
electric field. Then, the reflected electric field is given by 
\begin{eqnarray}
E_{R}^{x}(\omega ) &=&\frac{1-\sqrt{\varepsilon _{xx}}}{1+\sqrt{\varepsilon
_{xx}}}E_{0}(\omega ),  \label{fillo} \\
E_{R}^{y}(\omega ) &=&\frac{1-\sqrt{\varepsilon _{v}}}{1+\sqrt{\varepsilon
_{v}}}E_{0}(\omega ),  \notag
\end{eqnarray}%
with $\varepsilon _{v}=\varepsilon _{yy}+\varepsilon _{yz}^{2}/\varepsilon
_{zz}.$

\begin{figure}
\centering
\includegraphics[width = \linewidth]{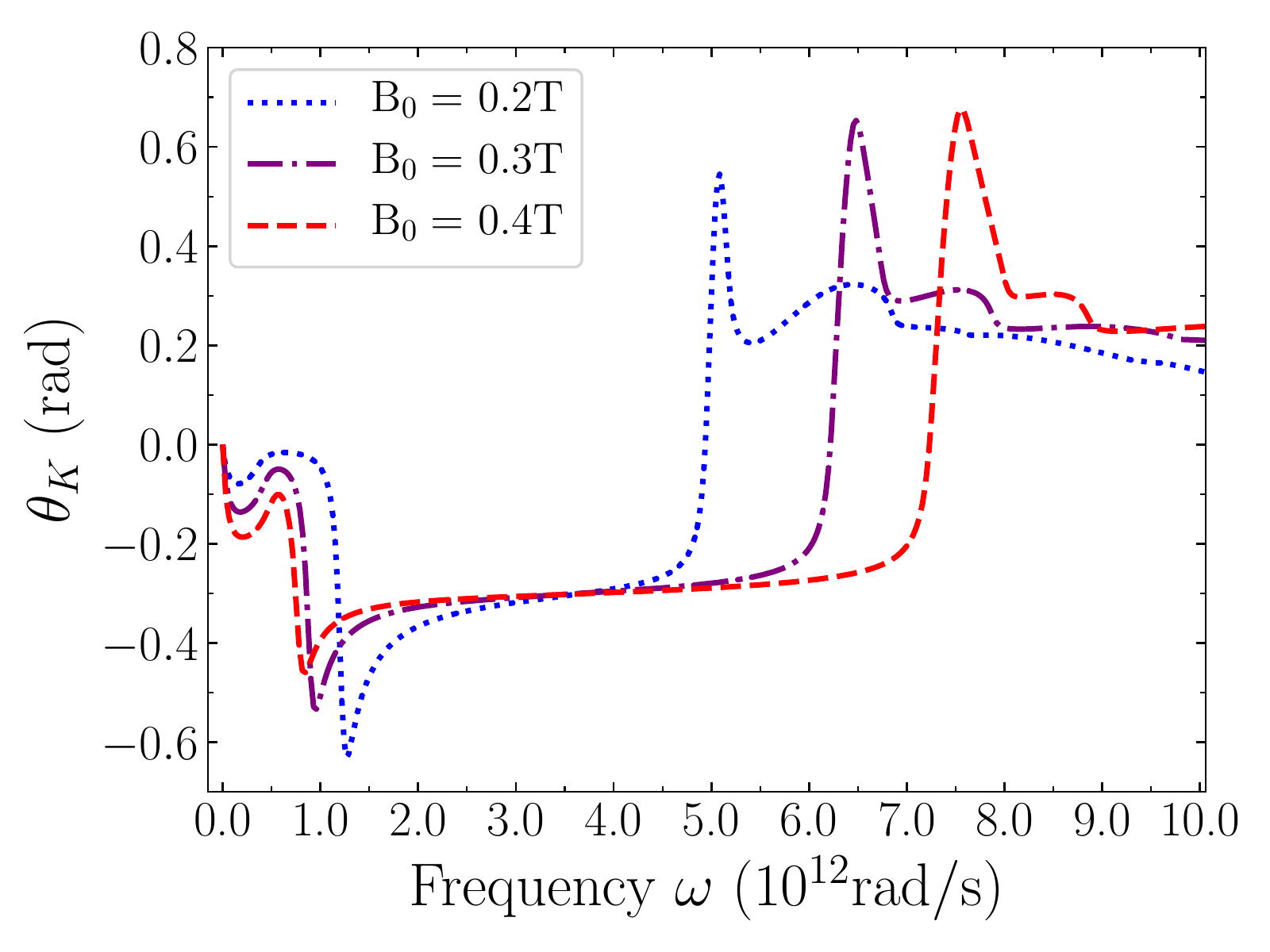} 
\caption{Kerr angle as a function of
		frequency for different magnetic fields in the Voigt configuration. 
		The blueshift of the large peak in $\protect\theta _{K}$ is related to the chiral anomaly.
		Parameters: $n_{e}=1\times 10^{20}$ m$^{-3},\protect\tau _{0}=10$ ps, $v_{F}=3\times
		10^{5}$ m/s, $\mathbf{t}=0.5\protect\widehat{\mathbf{z}}.$} \label{fig7}
\end{figure}

\begin{figure}
\centering\includegraphics[width = \linewidth]{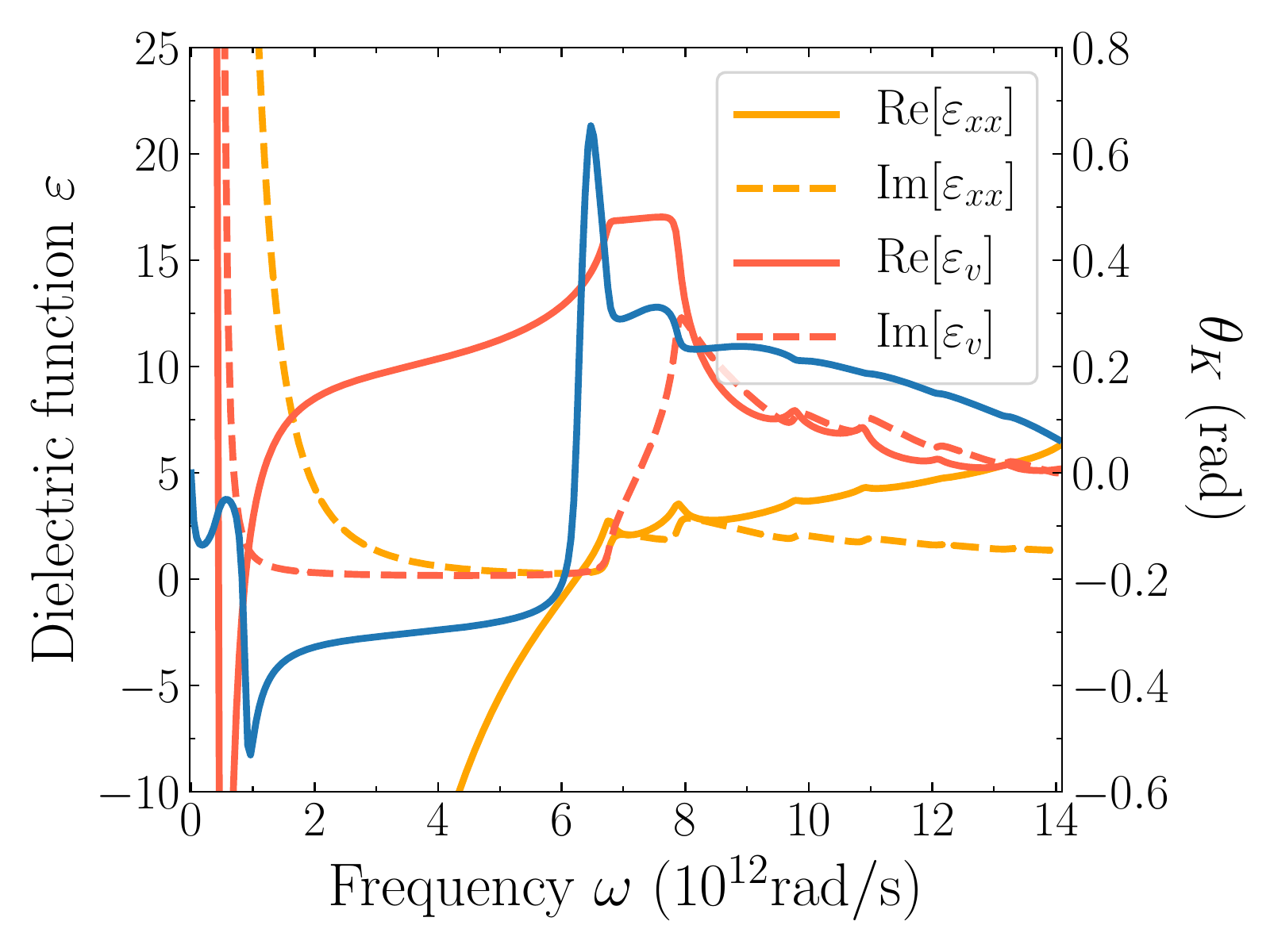} 
\caption{Frequency dependence of the
		dielectric functions $\protect\varepsilon _{xx}$ and $\protect\varepsilon _{v}
		$ and of the Kerr angle $\protect\theta _{K}$ for a magnetic field $B=0.3$
		T. Parameters: $n_{e}=1\times 10^{20}$ m$^{-3},\protect\tau _{0}=10$ ps, $		v_{F}=3\times 10^{5}$ m/s, $\mathbf{t}=0.5\protect\widehat{\mathbf{z}}.$ The
		first four peaks in $\func{Re}[\protect\varepsilon _{xx}]$ come from the
		transitions: $0\rightarrow 1,-1\rightarrow 0,0\rightarrow 2,-2\rightarrow 0.$}
\label{fig8}
\end{figure}

We first remark that if interband transitions are not included in the
calculation of the dielectric functions, then $\varepsilon _{yy}=1$ and $%
\varepsilon _{yz}=0$ so that $\varepsilon _{v}=1.$ It then follows from Eq. (%
\ref{fillo}) that $E_{R}^{y}(\omega )=0$, and so $\tan \left( \theta
_{K}+\theta _{I}\right) =0$, giving $\theta _{K}=-\pi /4.$ There is no
structure in the Kerr angle in this case. When both intraband and interband
transitions are included, the behavior of the Kerr angle with frequency and
magnetic field is shown in Fig. \ref{fig7}. We notice that the Kerr angle in
this configuration is as big as in the Faraday configuration, in contrast to
a normal metal (where it is at least ten times smaller). Moreover, its
behavior is clearly different from that in the Faraday configuration. The
strong peak in the Kerr angle at high frequency occurs when $\func{Re}\left[
\varepsilon _{xx}\left( \omega \right) \right] =0,$ i.e. at the WSM\ plasmon
frequency $\omega =\omega _{p}$ given by Eq. (\ref{plasmon}), as can be seen
in Fig. \ref{fig8}. The peak is also present at zero tilt, and its frequency
increases as $\sqrt{B_{0}}$. In an ordinary metal, the frequency of the peak
in the Kerr angle increases proportionally to $B_{0}^{2}$. Moreover, in a
metal, the negative peak in the Kerr angle occurs at the plasmon frequency $%
\omega _{p,metal}=\sqrt{ne^{2}/m\varepsilon _{0}}$, which is independent of $%
B_{0}$. In contrast, in a WSM, the negative peak occurs when $\func{Re}\left[
\varepsilon _{v}\left( \omega \right) \right] =0$ (see Fig. \ref{fig8}) and
is redshifted with magnetic field. In this figure, the plasmon frequency is
below the interband absorption threshold. This is a common circumstance,
especially if we take into account the effect of $\epsilon _{\infty }$ which
reduces the plasmon frequency without changing the optical gap for interband
transitions.

The blueshift of the Kerr angle in a WSM\ is consistent with the measurement
by Levy et al. (see Fig. 2(d) of Ref. [\onlinecite{Levy}]) of the blueshift
of the reflectance edge in the Weyl semimetal TaAs. As we pointed out in
Sec. III, this magnetoresistance effect is a signature of the chiral anomaly
in a WSM and, as Fig. \ref{fig7} clearly shows, it is reflected in the
displacement of the peak in the Kerr angle. This peak is caused mainly by
the intraband contribution of the chiral Landau level to the longitudinal
conductivity $\sigma _{xx}$ (i.e. $\sigma _{\Vert }$). Since it occurs at
the plasmon frequency given in Eq. (\ref{plasmon}), which is independent of $%
\tau _{0}$ in the limit $\omega _{p}\tau _{0}\gg 1,$ it is not affected by
the magnetic-field dependence of $\tau _{0}.$ We have checked numerically
that taking $\tau _{0}\propto 1/B_{0}$ or $\tau_0\propto 1/B_0^2$ does not
change the position of this peak.

The peak disappears if the intraband part of $\sigma _{xx}$ is set to zero,
but is almost unchanged if $\sigma _{zz}$ (i.e. $\sigma _{\bot }$) is zero.
As discussed in Sec. III, the interband transitions produce a redshift of
the plasmon frequency. The shift of the Kerr angle occurs in the THz range
and should be measurable in optical experiments. The blueshift of the Kerr
angle with magnetic field in the Voigt geometry contrasts with its behavior
in the Faraday configuration, where it is redshifted with the magnetic
field. The Voigt geometry is special because the longitudinal dielectric
function (e.g. $\varepsilon _{xx}$ when $\mathbf{B}_{0}||\hat{\mathbf{x}}$)
enters the definition of the Kerr angle, while it does not in the Faraday
configuration.

Aside from the large maximum caused by intraband transitions, the Kerr angle
displays additional smaller peaks originating from interband transitions.
The first four of these peaks, also visible in $\func{Re}\left[ \varepsilon
_{xx}\left( \omega \right) \right] $ (see Fig. \ref{fig8}), occur at the
transitions $0\rightarrow 1;$ $-1\rightarrow 0;0\rightarrow 2$ and $%
-2\rightarrow 0$. Because we choose $\mathbf{t}\bot \mathbf{B}$ (i.e. $%
t_{\Vert }=0$ in Eq. (\ref{goerg})), there is no splitting of the $%
0\rightarrow 1$ and $-1\rightarrow 0$ peaks and so no valley polarization in
the Kerr angle.

Figures \ref{fig7} and \ref{fig8} show some intriguing structure in the
low-frequency region. However, as this low-frequency behavior will likely be
strongly affected by details of the disorder, for example the impact of the $%
B-$dependence of the scattering time, we will not discuss it further in this
paper.

\section{SUMMARY}

We have shown that a sizeable Kerr rotation is possible in a simplified
model of a Weyl semimetal in both the Faraday and Voigt configurations.
Inter-Landau level transitions appear as dips or peaks in a plot of the Kerr
angle \textit{vs} frequency and can be detected in this way. Moreover, a
tilt of the Weyl cones allows the detection of transitions beyond the
dipolar ones. In the particular Faraday configuration that we have chosen to
study, with a nonzero tilt of the Weyl cones in the direction of the
external magnetic field, the peaks in the Kerr angle that represent
interband transitions involving the chiral Landau level are split into
subpeaks. This splitting, which we found before in the absorption spectrum,
is a manifestation of a full valley polarization in a WSM\cite{Bertrand2019}%
. A valley polarization is thus also detectable via the magneto-optical Kerr
effect.

We have also demonstrated the interest of measuring the MOKE\ in the Voigt
geometry. In this configuration, the Kerr angle involves the longitudinal
dielectric function $\varepsilon _{\Vert }\left( \omega \right) $, unlike in
the Faraday configuration, since the electric field of the wave has a
nonzero projection on the external magnetic field. The zero of the
longitudinal dielectric function gives the plasmon frequency $\omega _{p}$.
The chiral anomaly in a WSM leads to an increase of $\omega _{p}$ with the
external static magnetic, i.e. $\omega _{p}\propto \sqrt{B_{0}}$. In the
Voigt geometry, a large peak in the Kerr angle $\theta _{K}$ occurs at $%
\omega _{p}$ and is consequently blueshifted by a magnetic field. We argued
that the $\sqrt{B_{0}}$ shift of this peak is the signature of the chiral
anomaly in the MOKE, just like the shift of the absorption edge recently
reported in reflectance measurements\cite{Levy}.

The exact size of the Kerr angle depends, of course, on the dielectric
constant $\varepsilon _{\infty }$, which can be quite large in WSMs. We have
verified that using a larger value of $\varepsilon _{\infty }$ in our
calculation (instead of $\varepsilon _{\infty }=1$) does indeed reduce the
amplitude of the Kerr angle signal and redshifts the frequencies of the
peaks and dips. For $\varepsilon _{\infty }=30,$ the amplitude of the peak
in $\theta _{K}$ at $B=0.3$ T in Fig. 7 is decreased from $0.65$ rad to $%
0.41 $ rad and $\omega _{p}$ is reduced to $\omega _{p}\approx 2.5\times
10^{12}$ rad/s. If $\varepsilon _{\infty }$ is very large, the peak $\theta
_{K}\left( \omega _{p}\right) $ may be blurred by disorder. For the chiral
anomaly to be detected in the Kerr angle, the plasmon frequency must satisfy
the condition $\omega _{p}\tau _{0}\gg 1.$

We have chosen a total electronic density of $n_{e}=1\times 10^{20}$ m$^{-3}$
for our calculation, as we did not have a particular WSM in mind. At such a
small density, the quantum limit is reached at a very small value of $B_{0}$%
, i.e. $\approx 0.053$ T. To be certain that this choice does not affect the
main conclusions of our paper, and to get a quantum limit closer to recently
reported electronic density values\cite{Levy,chicheng,Grassano}, we have
recalculated Fig. 7 for a much bigger total density of $n_{e}=1\times
10^{22} $ m$^{-3}$ and a magnetic field of $B_{0}=7.83$ T (the value of $%
B_{0}$ needed to obtain the same Fermi level as in Fig. 7 for $B_{0}=0.3$
T). The qualitative aspects of the curve remain unchanged, with almost the
same amplitude for the peaks and dips, but with the peak at $\theta
_{K}\left( \omega _{p}\right) $ now shifted to $\approx 32\times 10^{12}$
rad/sec, i.e. $\approx 21$ meV. Thus, an increase in the carrier density
could compensate for the redshift created by a larger value of $\varepsilon
_{\infty }.$

Although our calculations have been carried out for a toy model, the main
predictions derived from them are applicable to real WSMs. For example, let
us consider TaAs, which displays two symmetry-inequivalent multiplets of
Weyl nodes (8 W1 nodes and 16 W2 nodes). If $\mathbf{B}_{0}$ and $\mathbf{q}$
are both parallel to the $c$-axis of the crystal (Faraday configuration),
one should see a peak-splitting due to valley polarization in the
contribution from the W2 nodes to the Kerr angle. No such splitting would be
present in the contribution from the W1 nodes, because these are not tilted
along the $c$ axis. In principle, one can distinguish the contributions of
the W1 and W2 nodes because their optical absorption thresholds differ by
about $10$ meV. Also, if $\mathbf{B}_{0}$ is along the $c$ axis of the
crystal but $\mathbf{q}$ is along the $a$ or $b$ axis (Voigt configuration),
a peak in the Kerr angle will emulate the blueshift of the plasmon frequency
as a function of $B_{0}$. Finally, the contribution from Fermi arcs to the
magneto-optical response, neglected in our theory, should not significantly
affect our main predictions, as the penetration depth of the electromagnetic
waves in weakly doped bulk WSMs can largely exceed the localization length
of the surface states.

\begin{acknowledgments}
R. C. and I. G. were supported by grants from the Natural Sciences and
Engineering Research Council of Canada (NSERC). J.-M. P. was supported by
scholarships from NSERC and FRQNT. Computer time was provided by Calcul Qu%
\'{e}bec and Compute Canada. The authors thank S. Bertrand for helpful
discussions.
\end{acknowledgments}

\appendix

\section{EIGENSTATES\ OF\ TILTED WEYL\ CONES\ }

A transverse static magnetic field $\mathbf{B}=B_{0}\widehat{\mathbf{z}}$ is
added via the Peierls substitution $\mathbf{p}\rightarrow \mathbf{P}=\mathbf{%
p}+e\mathbf{A,}$ with the vector potential $\mathbf{A}=\left( 0,Bx,0\right) $
($e>0$ for an electron) taken in the Landau gauge. The Hamiltonian $h_{1}(%
\mathbf{p})$ for node $1$ becomes 
\begin{equation}
h_{1}(\mathbf{P})=h_{1}^{\left( 0\right) }(\mathbf{P})+W_{1}\left( \mathbf{P}%
\right) ,
\end{equation}%
with 
\begin{equation}
h_{1}^{\left( 0\right) }(\mathbf{P})=v_{F}\left( 
\begin{array}{cc}
P_{z} & P_{x}-iP_{y} \\ 
P_{x}+iP_{y} & -P_{z}%
\end{array}%
\right) ,
\end{equation}%
the Hamiltonian in the absence of a tilt, and with%
\begin{equation}
W_{1}\left( \mathbf{P}\right) =v_{F}\left( 
\begin{array}{cc}
\mathbf{t\cdot P} & 0 \\ 
0 & \mathbf{t\cdot P}%
\end{array}%
\right),
\end{equation}%
the perturbation due to the tilt.

The Hamiltonian $h_{1}^{\left( 0\right) }(\mathbf{p})$ is easily
diagonalized by defining the ladder operators%
\begin{eqnarray}
a &=&\frac{\ell }{\sqrt{2}\hslash }\left( P_{x}-iP_{y}\right) , \\
a^{\dag } &=&\frac{\ell }{\sqrt{2}\hslash }\left( P_{x}+iP_{y}\right) .
\end{eqnarray}%
The energy spectrum of $h_{1}^{\left( 0\right) }(\mathbf{p})$ consists in
both positive and negative energy Landau levels. We classify them with the
indices $n,s$ where $n=0,1,2,3,...$ is always positive and $s=+1\left(
-1\right) $ is a band index for the positive(negative) energy levels. For
node $1:$%
\begin{eqnarray}
E_{1,0}\left( p_{z}\right) &=&-v_{F}p_{z}, \\
E_{1,n\neq 0,s}\left( p_{z}\right) &=&sv_{F}\sqrt{2\frac{\hslash ^{2}}{\ell
^{2}}n+p_{z}^{2}},
\end{eqnarray}%
with the corresponding eigenstates 
\begin{equation}
w_{1,n,s,X}\left( p_{z},\mathbf{r}\right) =\frac{1}{\sqrt{L_{z}}}\left( 
\begin{array}{c}
u_{1,n,s}\left( p_{z}\right) h_{n-1,X}\left( x,y\right) \\ 
v_{1,n,s}\left( p_{z}\right) h_{n,X}\left( x,y\right)%
\end{array}%
\right) e^{ip_{z}z/\hslash }.
\end{equation}%
In these equations, $\ell =\sqrt{\hslash /eB}$ is the magnetic length and
the wave functions $h_{n,X}\left( x,y\right) $ are the eigenstates of a
two-dimensional electron gas in a transverse magnetic field in the Landau
gauge: 
\begin{equation}
h_{n,X}\left( x,y\right) =\frac{1}{\sqrt{L_{y}}}\varphi _{n}\left(
x-X\right) e^{-iXy/\ell ^{2}},
\end{equation}%
with $\varphi _{n}\left( x\right) $ the eigenstates of the one-dimensional
harmonic oscillator and $n$ and $X$ the Landau level and guiding-center
indices respectively.

The coefficients $u_{1,n,s,p_{z}},v_{1,n,s,p_{z}},$ which are independent of 
$X,$ are given for $n\neq 0$ by 
\begin{equation}
\left( 
\begin{array}{c}
u_{1,n,s}\left( p_{z}\right) \\ 
v_{1,n,s}\left( p_{z}\right)%
\end{array}%
\right) =\frac{1}{\sqrt{2}}\left( 
\begin{array}{c}
-is\sqrt{1-\frac{v_{F}p_{z}}{E_{1,n,s}\left( p_{z}\right) }} \\ 
\sqrt{1+\frac{v_{F}p_{z}}{E_{1,n,s}\left( p_{z}\right) }}%
\end{array}%
\right) ,
\end{equation}%
and obey the normalization condition $\left\vert u_{1,n,s}\left(
p_{z}\right) \right\vert ^{2}+\left\vert v_{1,n,s}\left( p_{z}\right)
\right\vert ^{2}=1.$ For the chiral Landau level $n=0$, the eigenstate is
simply 
\begin{equation}
\left( 
\begin{array}{c}
u_{1,0}\left( p_{z}\right) \\ 
v_{1,0}\left( p_{z}\right)%
\end{array}%
\right) =\left( 
\begin{array}{c}
0 \\ 
1%
\end{array}%
\right)
\end{equation}%
and there is no band index in this case. The eigenenergies are independent
of $X$ so that each level $\left( n,s,p_{z}\right) $ has a degeneracy given
by $N_{\varphi }=S/2\pi \ell ^{2}$ with $S=L_{x}L_{y}$ the area of the
electron gas in the $x-y$ plane ($L_{z}$ is its extension in the $z$
direction).

We must now obtain the energies and eigenvectors for tilted cones. Although
an analytical solution exists\cite{Goerbig2016}, it is more convenient to
use a numerical approach to the eigenvalue problem since we can then study
the system for arbitrary orientations of the magnetic field and tilt vector
or consider more complex Hamiltonians with, for example, non linear terms in
the energy spectrum\cite{Bertrand2017}.

We write the many-body Hamiltonian for the four-node model as%
\begin{equation}
\mathcal{H}=\sum_{\tau }\int d^{3}r\psi _{\tau }^{\dag }\left( \mathbf{r}%
\right) h_{\tau }(\mathbf{r})\psi _{\tau }\left( \mathbf{r}\right)
\end{equation}%
and expand the fermionic field operators $\psi _{\tau }\left( \mathbf{r}%
\right) $ onto the $\left\{ w_{\tau ,n,s,X}\left( p_{z},\mathbf{r}\right)
\right\} $ basis (we leave as implicit the fact that $\mathbf{t}$ depends on
the node index $\tau $) 
\begin{equation}
\psi _{\tau }\left( \mathbf{r}\right) =\sum_{n,X,p_{z}}w_{\tau ,n,s,X}\left(
p_{z},\mathbf{r}\right) c_{\tau ,n,s,X,p_{z}},
\end{equation}%
where $c_{\tau ,n,s,X,p_{z}}$ is the annihilation operator for an electron
in the $w_{\tau ,n,s,X,p_{z}}\left( \mathbf{r}\right) $ state. The many-body
Hamiltonian can be written as%
\begin{eqnarray}
\mathcal{H} &=&\sum_{\tau ,n,s,p_{z}}\left( E_{\tau ,n,s}\left( p_{z}\right)
+v_{F}t_{z}p_{z}\right) c_{\tau ,n,s,X,p_{z}}^{\dag }c_{\tau ,n,s,X,p_{z}} \\
&&+\frac{\hslash v_{F}}{\sqrt{2}\ell }\sum_{\tau ,n,n^{\prime },s,s^{\prime
},p_{z}}W_{\tau ,n,s,n^{\prime },s^{\prime }}\left( p_{z}\right) c_{\tau
,n,s,X,p_{z}}^{\dag }c_{\tau ,n^{\prime },s^{\prime },X,p_{z}}.  \notag
\end{eqnarray}%
The matrix elements $W_{\tau ,n,s,n^{\prime },s^{\prime }}\left(
p_{z}\right) $ are defined by%
\begin{eqnarray}
W_{\tau ,n,s,n^{\prime },s^{\prime }}\left( p_{z}\right) &=&\int
d^{3}rw_{\tau ,n,s,X}^{\dag }\left( p_{z},\mathbf{r}\right) \\
&&\times \left( t_{-}a^{\dag }+t_{+}a\right) w_{\tau ,n^{\prime },s^{\prime
}X}\left( p_{z},\mathbf{r}\right) ,  \notag
\end{eqnarray}%
with the definition%
\begin{equation}
t_{\pm }=t_{x}\pm it_{y}.
\end{equation}%
Their calculation gives%
\begin{eqnarray}
&&W_{\tau ,n,s,n^{\prime },s^{\prime }}\left( p_{z}\right) \\
&=&-it_{+}u_{\tau ,n,s}^{\ast }\left( p_{z}\right) u_{\tau ,n+1,s^{\prime
}}\left( p_{z}\right) \sqrt{n}\delta _{n^{\prime },n+1}  \notag \\
&&-it_{+}v_{\tau ,n,s}^{\ast }\left( p_{z}\right) v_{\tau ,n+1,s^{\prime
}}\left( p_{z}\right) \sqrt{n+1}\delta _{n^{\prime },n+1}  \notag \\
&&+it_{-}u_{\tau ,n,s}^{\ast }\left( p_{z}\right) u_{\tau ,n-1,s^{\prime
}}\left( p_{z}\right) \sqrt{n-1}\delta _{n^{\prime },n-1}  \notag \\
&&+it_{-}v_{\tau ,n,s}^{\ast }\left( p_{z}\right) v_{\tau ,n-1,s^{\prime
}}\left( p_{z}\right) \sqrt{n}\delta _{n^{\prime },n-1}.  \notag
\end{eqnarray}%
The linear dispersion being valid only in a small region around each Weyl
node, an appropriate high-energy cutoff must be applied to the summation
over $p_{z}\mathbf{.}$ That cutoff is assumed to be small compared to the
internodal separation in momentum space.

We define the super-indices $I=n,s$ and $J=n^{\prime },s^{\prime }$ in order
to write the Hamiltonian $\mathcal{H}$ in the matrix form :%
\begin{equation}
\mathcal{H}=\sum_{\tau ,X,p_{z}}\sum_{I,J}c_{I}^{\dag }\left( \tau
,X,p_{z}\right) F_{I,J}\left( \tau ,p_{z}\right) c_{J}\left( \tau
,X,p_{z}\right) ,
\end{equation}%
where $c_{J}\left( \tau ,X,p_{z}\right) $ stands for the vector $\left(
c_{\tau ,1,X,p_{z},}c_{\tau ,2,X,p_{z},},c_{\tau ,3,X,p_{z},},...,c_{\tau
,N,X,p_{z},}\right) ,$ where $N$ is the number of Landau levels $n,s$ kept
in the calculation and $F_{I,J}\left( \tau ,p_{z}\right) $ is the $N\times N$
matrix given by 
\begin{equation}
F_{I,J}\left( \tau ,p_{z}\right) =\left( E_{\tau ,I}\left( p_{z}\right)
+v_{F}t_{z}p_{z}\right) \delta _{I,J}+W_{\tau ,I,J}\left( p_{z}\right) ,
\end{equation}%
which is independent of the guiding-center index $X.$

The matrix $\mathbf{F}$ is hermitian and so can be diagonalized by a unitary
transformation%
\begin{equation}
\mathbf{F}=\mathbf{UDU}^{\dag },
\end{equation}%
where $\mathbf{U}$ is the matrix of the eigenvectors of $F$ and $D$ the
diagonal matrix of its eigenvalues $E_{I}\left( \tau ,p_{z}\right) ,$ which
are the energy of the Landau levels in the presence of the tilt.

Defining new operators $d_{I}\left( \tau ,X,p_{z}\right) $ by 
\begin{equation}
d_{I}\left( \tau ,X,p_{z}\right) =\sum_{J}\left( U^{\dag }\right)
_{I,J}c_{J}\left( \tau ,X,p_{z}\right) ,
\end{equation}%
we get the final result%
\begin{equation}
\mathcal{H}=\sum_{\tau ,X,p_{z}}\sum_{I}E_{I}\left( \tau ,p_{z}\right)
d_{I}^{\dag }\left( \tau ,X,p_{z}\right) d_{I}\left( \tau ,X,p_{z}\right) .
\label{hamilton}
\end{equation}%
We also consider the case where the magnetic field is in the $\widehat{%
\mathbf{x}}$ direction. The analysis proceeds in a similar way.

The analytical results\cite{Goerbig2016} for the energy levels of the tilted
cone $\tau =1$ are%
\begin{eqnarray}
E_{n=0,p_{\Vert }} &=&v_{F}t_{\Vert }p_{\Vert }-\frac{v_{F}}{\gamma }%
p_{\Vert },  \label{goerg} \\
E_{n\neq 0,p_{\Vert }} &=&v_{F}t_{\Vert }p_{\Vert }+s\frac{1}{\gamma }\sqrt{%
v_{F}^{2}p_{\Vert }^{2}+\frac{2\hslash ^{2}nv_{F}^{2}}{\gamma \ell ^{2}}}, 
\notag
\end{eqnarray}%
\newline
where $p_{\Vert }=\mathbf{p}\cdot \widehat{\mathbf{B}}$, $t_{\Vert }=\mathbf{%
t}\cdot \widehat{\mathbf{B}}$ and $\gamma =\left( 1-\left\vert \mathbf{t}%
\times \widehat{\mathbf{B}}\right\vert ^{2}\right) ^{-1/2}.$

\section{CURRENT\ OPERATOR}

The single-particle current operator for node $\tau $ is defined by%
\begin{equation}
\mathbf{j}_{\tau }=-\left. \frac{\partial h_{\tau }(\mathbf{p})}{\partial 
\mathbf{A}_{ext}}\right\vert _{\mathbf{A}_{ext}\rightarrow 0}=-ev_{F}\left( 
\mathbf{\sigma }+\mathbf{t}\sigma _{0}\right) ,
\end{equation}%
where $\mathbf{A}_{ext}$ is the vector potential of an external
electromagnetic field. The total many-body current for this node is given by 
\begin{eqnarray}
\mathbf{J}_{\tau } &=&\int d^{3}r\psi _{\tau }^{\dag }\left( \mathbf{r}%
\right) \mathbf{j}_{\tau }\psi _{\tau }\left( \mathbf{r}\right) \\
&=&-ev_{F}\sum_{\tau ,p_{z},X}\sum_{I,J}\mathbf{\Lambda }_{I,J}\left( \tau
,p_{z}\right) c_{\tau ,I,X,p_{z}}^{\dag }c_{\tau ,J,X,p_{z}}  \notag \\
&=&-ev_{F}\sum_{\tau ,p_{z},X}\sum_{I,J}\mathbf{\Upsilon }_{I,J}\left( \tau
,p_{z}\right) d_{I}^{\dag }\left( \tau ,X,p_{z}\right) d_{J}\left( \tau
,X,p_{z}\right) ,  \notag
\end{eqnarray}%
with the matrix elements defined by%
\begin{equation}
\mathbf{\Lambda }_{I,J}\left( \tau ,p_{z}\right) =\int d^{3}rw_{\tau
,I,X,p_{z}}^{\dag }\left( \mathbf{r}\right) \left( \mathbf{\sigma }+\mathbf{t%
}\sigma _{0}\right) w_{\tau ,J,X,p_{z}}\left( \mathbf{r}\right)
\end{equation}%
and the definition%
\begin{equation}
\mathbf{\Upsilon }=\mathbf{U}^{\dag }\mathbf{\Lambda U.}  \label{matrixe}
\end{equation}%
The matrix elements%
\begin{eqnarray}
&&\mathbf{\Lambda }_{n,s,n^{\prime },s^{\prime }} \\
&=&u_{\tau ,n,s}^{\ast }\left( p_{z}\right) v_{\tau ,n-1,s^{\prime }}\left(
p_{z}\right) \delta _{n^{\prime },n-1}\left( \widehat{\mathbf{x}}-i\widehat{%
\mathbf{y}}\right)  \notag \\
&&+v_{\tau ,n,s}^{\ast }\left( p_{z}\right) u_{\tau ,n+1,s^{\prime }}\left(
p_{z}\right) \delta _{n^{\prime },n+1}\left( \widehat{\mathbf{x}}+i\widehat{%
\mathbf{y}}\right)  \notag \\
&&+u_{\tau ,n,s}^{\ast }\left( p_{z}\right) u_{\tau ,n,s^{\prime }}\left(
p_{z}\right) \delta _{n^{\prime },n}\widehat{\mathbf{z}}  \notag \\
&&-v_{\tau ,n,s}^{\ast }\left( p_{z}\right) v_{\tau ,n,s^{\prime }}\left(
p_{z}\right) \delta _{n^{\prime },n}\widehat{\mathbf{z}}  \notag \\
&&+\mathbf{t}\delta _{n,n^{\prime }}\delta _{s,s^{\prime }}.  \notag
\end{eqnarray}

\section{MOKE IN\ A\ NORMAL METAL}

For comparison with the Kerr rotation in a WSM, we give a short review of
the Kerr effect in a normal three-dimensional metal in a transverse magnetic
field. We consider a system in which the half-space $z<0$ (region 1)\ is
occupied by a non-conducting medium with relative dielectric constant $%
\varepsilon _{1}=1$ and the half-space $z>0$ (region 2) is occupied by the
normal metal. We use the results of Section IV, but with the relative
dielectric tensor $\overleftrightarrow{\mathbf{\varepsilon }}$ of a normal
metal.

\subsection{Longitudinal propagation $\left( \mathbf{q\parallel B}_{\mathbf{0%
}}\mathbf{\parallel \hat{z}}\right) $}

For a magnetic field $\mathbf{B}=B_{0}\mathbf{\hat{z},}$ the (relative)\
dielectric tensor for a normal metal in the Drude model has the same
symmetry as $\overleftrightarrow{\mathbf{\varepsilon }}$ in Eq. (\ref{tensor}%
). The components are given by%
\begin{eqnarray}
\varepsilon _{xx} &=&\varepsilon _{yy}=1-\frac{\omega _{p}^{2}}{\omega }%
\frac{\omega +\frac{i}{\tau _{m}}}{\left( \omega +\frac{i}{\tau _{m}}\right)
^{2}-\omega _{c}^{2}}, \\
\varepsilon _{xy} &=&-\varepsilon _{yx}=\frac{\omega _{p}^{2}}{\omega }\frac{%
i\omega _{c}}{\left( \omega +\frac{i}{\tau _{m}}\right) ^{2}-\omega _{c}^{2}}%
, \\
\varepsilon _{zz} &=&1-\frac{\omega _{p}^{2}}{\omega }\frac{1}{\omega +\frac{%
i}{\tau _{m}}},
\end{eqnarray}%
where $\tau _{m}$ is the transport relaxation time, $\omega _{c}=eB_{0}/m$
the cyclotron frequency and $\omega _{p}=\sqrt{ne^{2}/\varepsilon _{0}m}$,
with $\varepsilon _{0}$ the free space permittivity and $m$ the effective
mass of the electrons. The dispersion relation of Eq. (\ref{eq:q2_pm}) for
the two circularly polarized electromagnetic waves in a normal metal becomes%
\begin{equation}
q_{\pm }(\omega )=\frac{\omega }{c}\sqrt{\varepsilon _{\pm }},
\end{equation}%
with%
\begin{eqnarray}
\varepsilon _{\pm }\left( \omega \right) &=&\varepsilon _{xx}\mp
i\varepsilon _{xy}  \label{epsipm} \\
&=&1-\frac{\omega _{p}^{2}}{\omega }\left( \frac{1}{\omega \pm \omega _{c}+%
\frac{i}{\tau _{m}}}\right) .  \notag
\end{eqnarray}%
In the pure limit ($\tau _{m}\rightarrow \infty$), the zeros of $%
\varepsilon_\pm $ are given by (keeping only positive frequencies) 
\begin{equation}
\omega _{\pm }=\frac{\mp \omega _{c}+\sqrt{\omega _{c}^{2}+4\omega _{p}^{2}}%
}{2}.  \label{omegaplus}
\end{equation}%
We remark that Eq. (\ref{epsipm}) remains valid in the quantum case, when
the kinetic energy is quantized into Landau levels.

For an incident wave linearly polarized at $\theta _{I}=\pi /4$ from the $x$
axis with amplitude $E_{0}^{x}$, the reflection coefficients are given by
Eqs. (\ref{rxx})-(\ref{rxy}), the Kerr angle is obtained from Eq. (\ref%
{tanfara}) and the phases $\theta _{R}^{x}$ and $\theta _{R}^{y}$ are
defined in Eq. (\ref{phases}).

\begin{figure}
\centering\includegraphics[width = \linewidth]{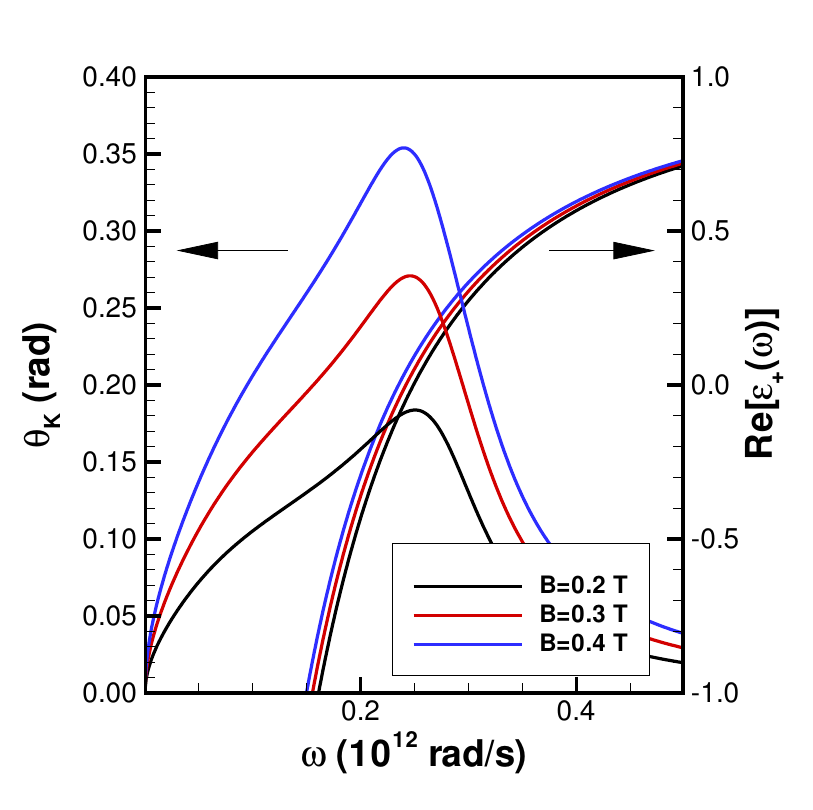} 
\caption{Kerr angle in the Faraday configuration (left $y$ axis) and dielectric
function  $\func{Re}[\varepsilon _{+}\left( \omega \right) ]$ (right $y$
axis) as a function of frequency for an external magnetic field $B_{0}=0.2,0.3,0.4$ and an electronic density $n_{e}=0.25\times 10^{20}$ m$^{-3}$.}
\label{fig9}
\end{figure}

Figure \ref{fig9} shows the Kerr angle $\theta _{K}$ (left $y$ axis)
calculated from Eq. (\ref{Standard}) (black line) for the same magnetic
fields $B_{0}=0.2,0.3,0.4$ T used in Fig. 7 and the same electronic density
(equivalent to the density of one node) $n_{e}=0.25\times 10^{20}$ m$^{-3}.$
The relaxation time is taken to be $\tau _{m}=10$ ps. As in a WSM, the big
peak in the Kerr angle occurs near the zero of the dielectric function $%
\func{Re}\left[ \varepsilon _{+}\left( \omega \right) \right] $ (right $y$
axis), i.e. at $\omega _{+},$ and is shifted to lower frequencies with
increasing magnetic field. There is no particular feature of the Kerr angle
at the cyclotron frequency $\omega _{c}=eB_{0}/m,$ which is $\omega
_{c}=0.035\times 10^{12}$ rad/s for $B=0.2$ T. The density that we use in
these calculations is much smaller than that in a normal metal. Our goal is
to compare the predictions of the Drude model with that of the WSM, where
our numerical calculation uses that particular density.

Figure 10 shows what happens if we increase the density to $n_{e}=0.25\times
10^{22}$ m$^{-3}.$ There is more structure in the Kerr angle, but the
maximum still occurs at $\func{Re}[\varepsilon _{+}\left( \omega _{+}\right)
]=0,$ i.e. at a frequency $10$ times bigger, and is again redshifted in
frequency by an increasing magnetic field. The shoulder on the left occurs
at the smallest of the two zeros of $\func{Re}[\varepsilon _{-}\left( \omega
\right) ].$ At a density of $n_{e}=1\times 10^{29}$ m$^{-3},$ typical of a
real metal and keeping $\tau _{m}=10$ ps with $B_{0}=10$ T, the Kerr angle
is almost constant at $\theta _{K}=0.1$ mrad. It peaks at $\func{Re}%
[\varepsilon _{+}\left( \omega _{+}\right) ]=0,$ i.e. at $\omega
_{+}=18.9\times 10^{15}$ rad/s where it reaches $3.5$ mrad. The frequency $%
\omega _{+}$ is then given by Eq. (\ref{omegaplus}), with $\omega _{c}\tau
_{m}\ll\omega _{p}\tau _{m}$ and $\omega _{+}=\omega _{p}.$ The Kerr angle
decreases with decreasing value of $\tau _{m}.$

\begin{figure}
\centering\includegraphics[width = \linewidth]{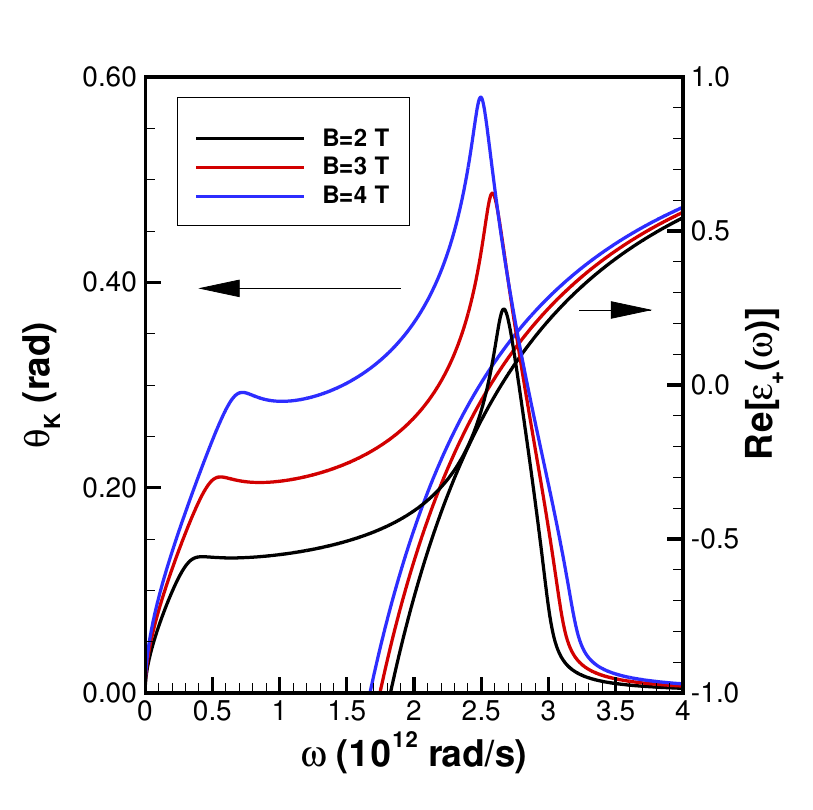} 
\caption{Kerr angle in the Faraday configuration (left $y$ axis) and dielectric
function $\func{Re}[\varepsilon _{+}\left( \omega \right) ]$ (right $y$
axis) as a function of frequency for an external magnetic field $B_{0}=2,3,4$
and an electronic density $n_{e}=0.25\times 10^{22}$ m$^{-3}$} \label{fig10}
\end{figure}

\subsection{Transverse propagation $\left( \mathbf{q\perp \mathbf{B}}_{0}%
\mathbf{\parallel \hat{x}}\right) $}

With $\mathbf{B}=B_{0}\widehat{\mathbf{x}},$ the dielectric tensor for a
normal metal has the symmetry of $\overleftrightarrow{\mathbf{\varepsilon }}$
in Eq. (\ref{tensor2}), except that $\varepsilon _{yy}=\varepsilon _{zz}$ and

\begin{eqnarray}
\varepsilon _{xx} &=&1-\frac{\omega _{p}^{2}}{\omega }\frac{1}{\left( \omega
+\frac{i}{\tau _{m}}\right) }, \\
\varepsilon _{yy} &=&\varepsilon _{zz}=1-\frac{\omega _{p}^{2}}{\omega }%
\frac{\omega +\frac{i}{\tau _{m}}}{\left( \omega +\frac{i}{\tau _{m}}\right)
^{2}-\omega _{c}^{2}}, \\
\varepsilon _{yz} &=&-\varepsilon _{zy}=-\frac{\omega _{p}^{2}}{\omega }%
\frac{i\omega _{c}}{\left( \omega +\frac{i}{\tau _{m}}\right) ^{2}-\omega
_{c}^{2}}.
\end{eqnarray}

Maxwell equations for this Voigt configuration give two solutions with the
polarization parallel or perpendicular to the magnetic field and dispersion%
\begin{eqnarray}
q_{\parallel } &=&\frac{\omega }{c}\sqrt{\varepsilon _{xx}}, \\
q_{\perp } &=&\frac{\omega }{c}\sqrt{\varepsilon _{v}},
\end{eqnarray}%
where the Voigt dielectric constant is defined by 
\begin{equation}
\varepsilon _{v}=\varepsilon _{yy}+\frac{\varepsilon _{yz}^{2}}{\varepsilon
_{zz}}.
\end{equation}%
In this configuration there is an induced field in the direction of
propagation, which is given by%
\begin{equation}
E_{z}=-\frac{\varepsilon _{zy}}{\varepsilon _{zz}}E_{y}.
\end{equation}

To see any variations of the Kerr angle, the incident wave has to be
polarized at an angle with respect to the $x$ and $y$ axis. The reflected
electric field components are then obtained from Eq. (\ref{fields}), while
the Kerr angle is obtained from Eq. (\ref{tanvoigt}). In our calculation, we
assume that the incident wave is linearly polarized at an angle $\theta
_{I}=\pi /4$ with respect to the $x$ axis, so that $E_{0}^{x}(\omega
)=E_{0}^{y}(\omega ).$ The dielectric function $\func{Re}\left[ \varepsilon
_{xx}\left( \omega \right) \right] =0$ at $\omega =\omega _{p}=\sqrt{%
ne^{2}/m\varepsilon _{0}}.$

\begin{figure}[H]
\centering\includegraphics[width = \linewidth]{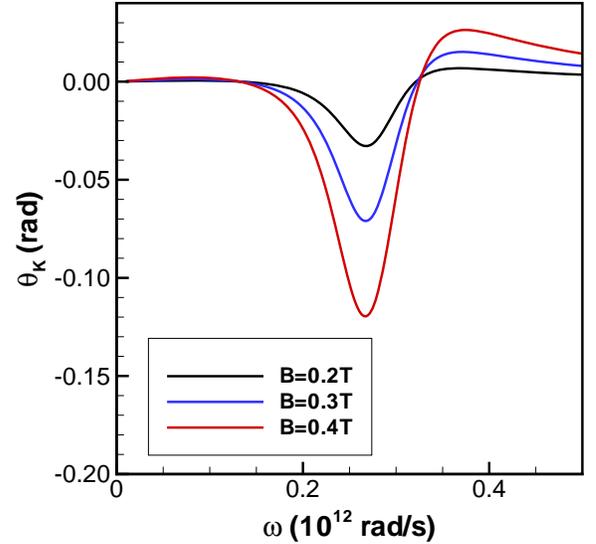} 
\caption{Kerr angle $\protect\theta _{K}$ in the Voigt configuration 
		as a function of frequency for $B_{0}=0.2,0.3,0.4$ T. 
		Parameters: $n_{e}=0.25\times 10^{20}$ m$^{-3}$ and $\protect\tau _{m}=10$ ps.}
\label{fig11}
\end{figure}

Figure \ref{fig11} shows the Kerr angle $\theta _{K}$ in the Voigt
configuration for three different values of the magnetic field, i.e. $%
B_{0}=0.2,0.3,0.4$ T, the same values as in Fig. 7 for the WSM. The carrier
density is $n_{e}=0.25\times 10^{20}$ m$^{-3}$ and the relaxation time is $%
\tau _{m}=10$ ps. The Kerr angle is an order of magnitude smaller than in
the Faraday configuration. Its minimum occurs near the plasmon frequency
while, in a WSM (see Fig. 7), it is the maximum of the Kerr angle that
occurs at that frequency. Because the plasmon frequency is independent of $%
B_{0}$ in a metal, the frequency at which the minimum in $\theta _{K}$
occurs is also independent of $B_{0}.$ The minimum in $\theta _{K}$
increases negatively with $B_{0}.$ We find numerically that the frequency at
which the Kerr angle is maximal $\theta _{K,\max }\propto B_{0}^{2}$ while $%
\theta _{K,\max }\propto \sqrt{B_{0}}$ in a WSM. In contrast with a WSM,
this maximum does not seem to be associated with any feature in $\varepsilon
_{v}$ or $\varepsilon _{xx}.$

\end{document}